\documentclass[12pt]{article}

\usepackage{geometry, fullpage, authblk}
\usepackage{xr-hyper}
\usepackage{times}
\usepackage{hyperref}[]
\hypersetup{
	colorlinks=true,
	linkcolor=darkblue,
	filecolor=magenta,      
	urlcolor=darkblue,
	citecolor=blue,
}
\usepackage[normalem]{ulem}
\usepackage{multicol}

\usepackage{url}

\usepackage{bigints}
\usepackage{amsfonts,amsmath,amssymb,amsthm, bbm}
\usepackage{wrapfig}
\usepackage{verbatim,float,url}
\usepackage{graphicx}
\usepackage{sidecap}
\usepackage{subfigure}
\usepackage[skip=0pt,font=footnotesize]{subcaption}
\usepackage[skip=0pt,font=footnotesize]{caption}
\usepackage{natbib}
\usepackage[english]{babel}
\usepackage{cancel}

\usepackage{color}
\definecolor{darkblue}{RGB}{10, 10, 200}

\usepackage[dvipsnames]{xcolor}
\usepackage{cleveref}
\usepackage{enumerate}
\usepackage{multirow}
\usepackage{amsmath}

\usepackage{collcell}
\makeatletter
\newcolumntype{G}{>{\collectcell\@gobble}c<{\endcollectcell}@{}}
\makeatother

\usepackage{bm}

\usepackage{mathtools}

\usepackage[dvipsnames]{xcolor}

\newtheorem{theorem}{Theorem}

\newtheorem{corollary}{Corollary}
\newtheorem{proposition}{Proposition}

\usepackage{mdwlist}

\usepackage{paralist}
\renewenvironment{itemize}[1]{\begin{compactitem}#1}{\end{compactitem}}
\renewenvironment{enumerate}[1]{\begin{compactenum}#1}{\end{compactenum}}

\theoremstyle{definition}
\newtheorem{definition}{Definition}

\usepackage{tikz}  
\usetikzlibrary{trees}
\usetikzlibrary{arrows.meta}
\usetikzlibrary{decorations.pathreplacing}
\usetikzlibrary{bayesnet}

\usepackage{ctable}
\usepackage{colortbl}
\usepackage{dsfont}

\usepackage{mathtools}

\newcommand{\eps}{\varepsilon}

\usepackage{ulem}

\usepackage[usenames,dvipsnames]{xcolor}

\usepackage[ruled,vlined]{algorithm2e}
\SetKwInput{kwInit}{Initialize}

\theoremstyle{plain}
\newtheorem{assumption}{Assumption}
\allowdisplaybreaks
\usepackage{bm}

\title{Scalable and Robust Spatial Prediction via Multi-Resolution Ensembles of Predictive Processes}

\newcommand{\abs}{
Gaussian processes provide a flexible framework for spatial prediction, but their computational cost limits applicability to large-scale data with large sample size $n$. Predictive processes (PPs), a popular low-rank approximation, mitigate this burden by projecting the original process onto a reduced set of $m\ll n$ inducing points. However, existing theory requires  $m$ to grow with $n$, creating a trade-off between accuracy and computational efficiency. We address this challenge by introducing an ensemble of PPs based on spatial partitioning, and propose a novel partitioning and patching scheme with desirable properties. By generalizing the convergence results of PPs, it becomes possible to explicitly balance scalability and accuracy: increasing the number of ensemble components slows down the convergence but substantially improves computational efficiency. We further show theoretically that, despite the limited approximation accuracy of PPs with fixed, small $m$, they provide predictions that are more robust to training data contamination. Motivated by these findings, we finally introduce a multi-resolution ensemble that combines ensembles over possibly overlapping coarse to fine partitions. Unlike previous multi-resolution approaches, our method targets predictive robustness rather than accurate covariance approximation. Simulations and large-scale geostatistical applications demonstrate that our approach delivers accurate, robust predictions while being  computationally efficient, thereby providing a practical and broadly applicable solution for spatial prediction.
\vspace{0.1in}

\noindent\textbf{Keywords:} Divide-and-conquer; Gaussian process;  low-rank approximation; spatial partitioning. 
}

\author[1]{Nicolas Bianco
}
\author[1]{Nadja Klein
}
\affil[1]{Scientific Computing Center, Karlsruhe Institute of Technology, Germany}

\usepackage{titlesec} 

\begin{document}
	
\maketitle
\thispagestyle{empty}

\centerline{\bf Abstract}
\medskip
\normalsize
\setlength{\parskip}{.2cm }
\setlength{\parindent}{0.5cm}
\abs

\clearpage
\pagenumbering{arabic}
\titlespacing*\paragraph{0pt}{4pt plus 4pt minus 1pt}{6pt plus 0pt minus 1pt}

\section{Introduction}
\label{sec:intro}
Gaussian Processes \citep[GPs;][]{RasWil2006} are widely used in statistics and machine learning for tasks such as regression, classification, Bayesian optimization, time series analysis, and spatial analysis. 

In this paper, we focus on spatial prediction based on a GP regression which assumes that observations $y=(y(s_1),\ldots,y(s_n))^\top$ at locations $s_i\in\mathcal{S}\subset\mathbb{R}^2$ arise from
\begin{equation}\label{eq:gp}
    y(s) = w(s) + \varepsilon(s), \qquad s \in \mathcal{S},
\end{equation} 
where  $w(s)$ is a zero-mean GP with covariance  $C_{nn}(\theta)=\{c(s_i,s_j;\theta)\}_{i,j=1}^n$, parameterized by $\theta=(\eta^2,\phi,\nu)^\top$, namely the marginal variance $\eta^2$, the range $\phi$, and Mat\'{e}rn smoothness parameter $\nu$. The term $\varepsilon(s)\sim N(0,\tau^2)$ represents independent Gaussian noise (nugget effect). 

Despite their popularity, GPs scale poorly with $n$ because inference requires inversion of $C_{nn}(\theta)+\tau^2I_n$, incurring $O(n^3)$ complexity. This bottleneck severely limits their use in large-scale applications.
To overcome this challenge, numerous scalable GP methods have been proposed, broadly categorized into global and local approximations \citep{LiuOngShe2020,VakJos2024}. 

Global methods approximate a single process over the whole domain. Low-rank approaches, such as predictive processes \citep[PPs;][]{BanGelFinSan2008} and sparse GPs \citep{TitHen2009}, summarize the data using a small set of representative points. Sparse approaches rely on covariance tapering \citep{FurGenNyc2006}, block structures \citep{SanJunHua2011, BuiTit2025}, or a sparse precision matrix, as in Vecchia approximations \citep{KatGui2021}, nearest-neighbor GPs \citep{DatBanFin2016}, and meshed processes \citep{PerDun2024}. Global methods capture broad trends, but often oversmooth local features. 

Local methods partition the domain and fit separate GPs within each region, providing adaptability and enabling parallelization, at the cost of potential discontinuities and aggregation artifacts \citep{SzaHadVan2025}. Recent developments in local approximation approaches include mixture-of-experts models \citep{EtiLawWad2024} and distributed GP frameworks, notably the Bayesian Committee Machine \citep{Tre2000}. The local approximate GP (LAGP) of \cite{GraApl2015} circumvents domain partitioning and focuses on prediction by constructing, for each new location, a surrogate GP based on a local subset of the data. More recent developments combine local expert models with deep neural gating networks, enhancing flexibility and scalability \citep{EtiLawWad2024}. While local approximations are well-suited for spatially heterogeneous processes due to their region-specific adaptability, their dependence on localized data can make them prone to undersmoothing and overfitting.

Another line of work combines inference from multiple models. Meta-kriging \citep{GuhBan2018} fits a full GP on small random subsets of the data and aggregates the resulting inference. Predictive stacking \citep{ZhaTanBan2025} fits GPs with different fixed covariance parameters on the entire dataset and averages their predictions with cross-validated weights. \cite{PreBan2026} combines the two by performing predictive stacking on random subsets drawn from the whole domain.

In summary, local approaches tend to be highly accurate when there are no outliers and the model is well specified, owing to their local (e.g., region-specific) adaptability. However, theoretical guarantees for their predictive accuracy remain limited \citep{SzaHadVan2025}. In contrast, global methods are effective at capturing broad trends and naturally mitigate the influence of local noise or outliers, making them attractive  for improving predictive robustness. Theoretically, global methods have been studied, for example, by \citet{SzaZhu2026} for Vecchia approximations, or by \citet{BurRasWil2020} and \citet{NieSza2022} for inducing variables.  \citet{SzaZan2019}  show that a global distributed GP approximation can perform poorly. However, to the best of our knowledge, their predictive robustness properties  have not been formally established in prior work.

In practice, outliers are often unknown and hence, it is difficult to decide \emph{a priori} whether predictive accuracy or robustness is more critical for a given dataset. Balancing these objectives, together with scalability, remains challenging and motivates the development of hybrid methodologies that can automatically adapt by combining the strengths of global and local approaches. This paper addresses this challenge by making three key contributions:
\begin{enumerate}
    \item We propose an Ensemble of Predictive Processes (EPP), a hybrid strategy that combines global low-rank approximations with local adaptivity via spatial partitioning. We introduce a novel partitioning scheme designed to ensure space-filling and balanced point densities across subregions, a fundamental property underlying our theoretical results. To maintain continuity at partition boundaries, which represents an important empirical feature, we develop a patching strategy that smoothly aggregates local predictions. We establish convergence rates for the ensemble predictive mean, quantifying the trade-off between statistical efficiency and computational cost. This contributes to the growing literature on theoretical guarantees for local approximations, which remain limited.
    \item We define training data contamination and investigate the predictive robustness of PPs and EPPs for finite sample sizes $n$.  To this end, we theoretically characterize the generalization error in this setting,  and obtain an optimal choice of the number of inducing points $m$ to improve their predictive performance.
    \item To balance both predictive  accuracy and robustness, we introduce a Multi-Resolution Ensemble of Predictive Processes (MREPP), which constructs ensembles over possibly overlapping coarse-to-fine partitions, which we call resolutions. Predictions are aggregated not only within spatial partitions  but also across partitions through resolutions. We show theoretically that the MREPP has predictive performance that is at least as good as any of its EPP components. MREPP differs from previous multi-resolution GP approximations \citep{NycBanHam2015,Kat2017} since it targets predictive robustness rather than covariance approximation and does not require nested partitions.
\end{enumerate}

\begin{figure}[ht]
    \centering
    \includegraphics[width=1\linewidth]{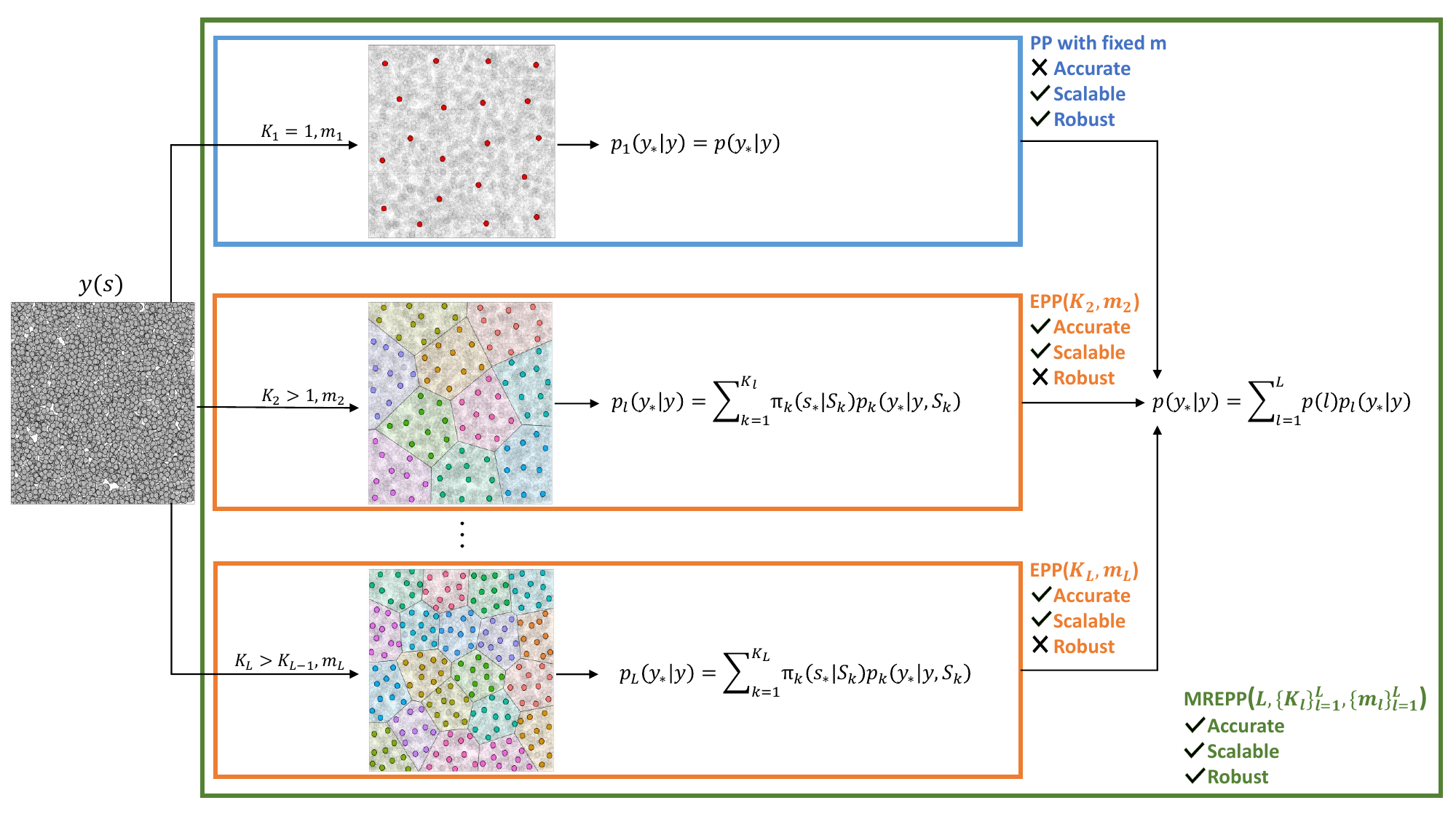}
    \caption{Graphical illustration for PP with fixed $m$ (blue box), the EPP (orange boxes for two distinct spatial partitions), and the MREPP (ensemble over EPPs of coarse-to-fine partitions in green box) and their properties investigated in this paper. The PP consists of a single resolution ($L=1$), and a single ensemble member ($K_1=1$). The spatial domain is approximated based on a fixed number $m_1=m$ of inducing points. 
    While the PP is scalable and more robust it has limited prediction accuracy. The EPP in contrast partitions the space into $K$ subregions, where each subregion $k=1,\ldots, K$ forms an ensemble member based on $m$ inducing points each.  This improves prediction accuracy compared to the PP at the price of a lack of robustness. As a solution, we propose MREPP, that leverages both the PP and EPP by combining inference over $L>1$ resolutions, with a coarse-to-fine structure of spatial domain partitions, where $K_L>K_{L-1}>\ldots>K_1$. At each resolution $l$ for $l=1,\ldots,L$ a corresponding EPP on $K_l$ subregions is fitted. Computations can be parallelized, yielding an approach that is simultaneously accurate, robust, and scalable.} 
    \label{fig:methods_graph}
\end{figure}

Figure \ref{fig:methods_graph} illustrates the general idea of the MREPP (green box) as a weighted ensemble over $L$ EPPs on coarse-to-fine partitions (orange boxes, depicted for two distinct spatial partitions). Each EPP consists of $K_l$, $l=1,\ldots,L$, PPs (blue box, shown for a single spatial partition). Data-driven weights  regulate the respective contributions  across $l=1,\ldots L$ resolutions and ensemble members  $k=1,\ldots, K_l$ at each resolution $l$ to balance predictive robustness and accuracy. Robustness arises when greater weight is assigned to coarser (i.e., early) resolutions, since these imply a smaller number of ensembles of PPs with fixed, small $m$ that are well-suited to estimate global trends and to reduce the influence of local outliers. In contrast,  predictive accuracy is preserved at finer (i.e., later) resolutions, where larger numbers of ensemble members and inducing points $m$ allow the model to better capture local behaviors.

The remainder of the paper is organized as follows: Section \ref{sec:PP} introduces the notation and reviews properties and limitations of PPs. Section \ref{sec:EPP} proposes the EPP and theoretically analyzes its predictive performance. Section \ref{sec:MREPP} presents the robustness properties of PPs and EPPs with a fixed number of inducing points and presents a novel multi-resolution ensemble methodology and its theoretical properties. Section \ref{sec:num} evaluates performance on simulated and real datasets. Section \ref{sec:concl} concludes and discusses future directions.

\section{Spatial predictive processes}
\label{sec:PP}
A predictive process (PP) provides a global low-rank approximation to the GP by projecting the original process onto a smaller set $\lbrace\tilde{s}_i\rbrace_{i=1}^m$ of $m$ inducing points, with $m\ll n$. Specifically, the  GP $w(s)$ in \eqref{eq:gp} is replaced by
\begin{equation}
\tilde{w}(s) = c_m(s,\theta) C_{mm}^{-1}(\theta) w_m,
\end{equation}
where $w_m \sim N_m(0,C_{mm}(\theta))$ is the GP evaluated at the inducing points, $c_m(s,\theta)$ is the $1\times m$ covariance vector between $s$ and the inducing points, and $C_{mm}(\theta)$ is the $m\times m$ covariance matrix between inducing points. This construction yields a rank-$m$ approximation to the full covariance matrix, reducing computational complexity from $O(n^3)$ to $O(nm^2)$ via the Woodbury identity.

Predictions at new locations follow from the joint Gaussian distribution obtained by integrating out $w_m$. Closed-form expressions for the predictive mean and covariance matrix  only involve $m\times m$ matrix inversions, and matrix-vector products with $n\times m$ cross-covariance terms, making PP a practical choice for large $n$. To keep the exposition concise, we defer  explicit formulas and  algebraic details to Supplementary Material A.

\subsection{Choosing the inducing points in predictive processes}
\label{sec:inducing_points}
A key component in PPs is the selection of inducing points, which governs both approximation accuracy and theoretical guarantees. Support Points (SPs) have emerged as a principled choice, offering space-filling coverage and ensuring convergence properties critical for PP approximations \citep{SonDaiGen2025}. SPs currently represent the state-of-the-art approach for inducing point selection. To analyze convergence, we next summarize mild regularity conditions, which are taken from \citet{SonDaiGen2025}, and which will  be assumed   in Sections~\ref{sec:EPP}--\ref{sec:MREPP}.

\begin{assumption}[Regularity conditions  \citep{SonDaiGen2025}]
\label{ass:regularity}
The observed locations $\{s_i\}_{i=1}^n$ are realizations of a random variable $s\sim F(s)$, where $F(s)$ is a finite Borel measure on $\mathcal{S}$. The Mat\'{e}rn covariance function  admits the Mercer expansion $c(s_i,s_j;\theta)= \sum_{k=0}^\infty  \lambda_k \phi_k(s_i)\phi_k(s_j)$ where $\lambda_k$ are positive eigenvalues such that $\lambda_1\geq \lambda_2 \geq \ldots$ and $\phi_k$ are the corresponding eigenfunctions. Assume the following hold:
\begin{enumerate}[(A)]
    \item $\mathcal{S}$ is a compact set with a Lipschitz boundary and satisfies an interior cone condition.
    \item The eigenvalues decay polynomially: $\lambda_i \leq \beta i^{-\gamma}$ for constants $\beta > 0$ and $\gamma > 1$.
    \item For any $i,j$,  $\int_{\mathcal{S}} \phi_i(s)^2 \phi_j(s)^2 \, dF(s)$ is bounded.  Moreover, the $3/2$-order derivatives of $\phi_i \phi_j$ are square-integrable on $\mathcal{S}$.
\end{enumerate}
\end{assumption}

These assumptions are common in the literature and \cite{SonDaiGen2025} provide a more detailed discussion on their practical implications. Under Assumption \ref{ass:regularity}, the authors study the convergence rate of the predictive means of the full GP, $\mu$, and of the PP, $\mu_m$, to the true mean function $\mu_0$. The latter is considered a random function with the same GP  as $w$. The convergence rates are quantified via the squared $L^2(F)$ norm $\| \mu_m - \mu_0 \|_{L^2(F)}^2 = \int_{\mathcal{S}}\big(\mu_m(s)-\mu_0(s)\big)^2 \, dF(s)$ and given by 
\begin{equation*}
\label{eq:fullgp_rate}
\| \mu - \mu_0 \|_{L^2(F)}^2 = O_p\big(n^{-(1 - 1/\gamma)}\big), \qquad n \to \infty,\ \gamma > 1
\end{equation*}
for the full GP. This coincides with the minimax rate for estimating a $2$-dimensional random function $\mu_0$ with smoothness $\gamma$ in nonparametric models \citep{VanVan2008,RosBorTer2023}. 
For the PP, the rate depends on the choice of inducing locations through $F$. Let $\hat{F}_m$ be the empirical measure of the $m$ inducing points and let $\mathcal{E}(F,\hat{F}_m)$\footnote{See \cite{SzeRiz2013}; $\mathcal{E}(F,\hat F_m)=\frac{2}{m}\sum_{i=1}^m \mathbb{E}\|s-\tilde{s}_i\|-\mathbb{E}\|s-s^\prime\|-\frac{1}{m^2}\sum_{i=1}^m\sum_{j=1}^m \|\tilde{s}_i-\tilde{s}_j\|$ where $s,s^\prime\sim F$ and $\{\tilde{s}_i\}_{i=1}^m$ inducing points.} denote their energy distance, which measures the discrepancy between the two distributions. If
\[
n^{2/\gamma}\,\mathcal{E}(F,\hat{F}_m) \to 0\, \text{ as } n,m \to \infty,\ \text{ and } \gamma > 2,
\]
then the PP predictive mean achieves the same minimax rate as the full GP:
\begin{equation}
\label{eq:pp_rate}
\| \mu_m - \mu_0 \|_{L^2(F)}^2 = O_p\big(n^{-(1 - 1/\gamma)}\big).
\end{equation}
In particular, SPs are optimal in the sense that they provide the best rate at which $\mathcal{E}(F,\hat{F}_m)$ approaches zero, namely $\mathcal{E}(F,\hat{F}_m) = O(1/m)$, implying the requirement $m \gtrsim n^{2/\gamma}$ of SPs to achieve the minimax rate \eqref{eq:pp_rate}.

\subsection{Limitations of PPs with SPs}
\label{sec:limitations}
\begin{figure}[ht]
    \centering
    \includegraphics[width=0.7\linewidth]{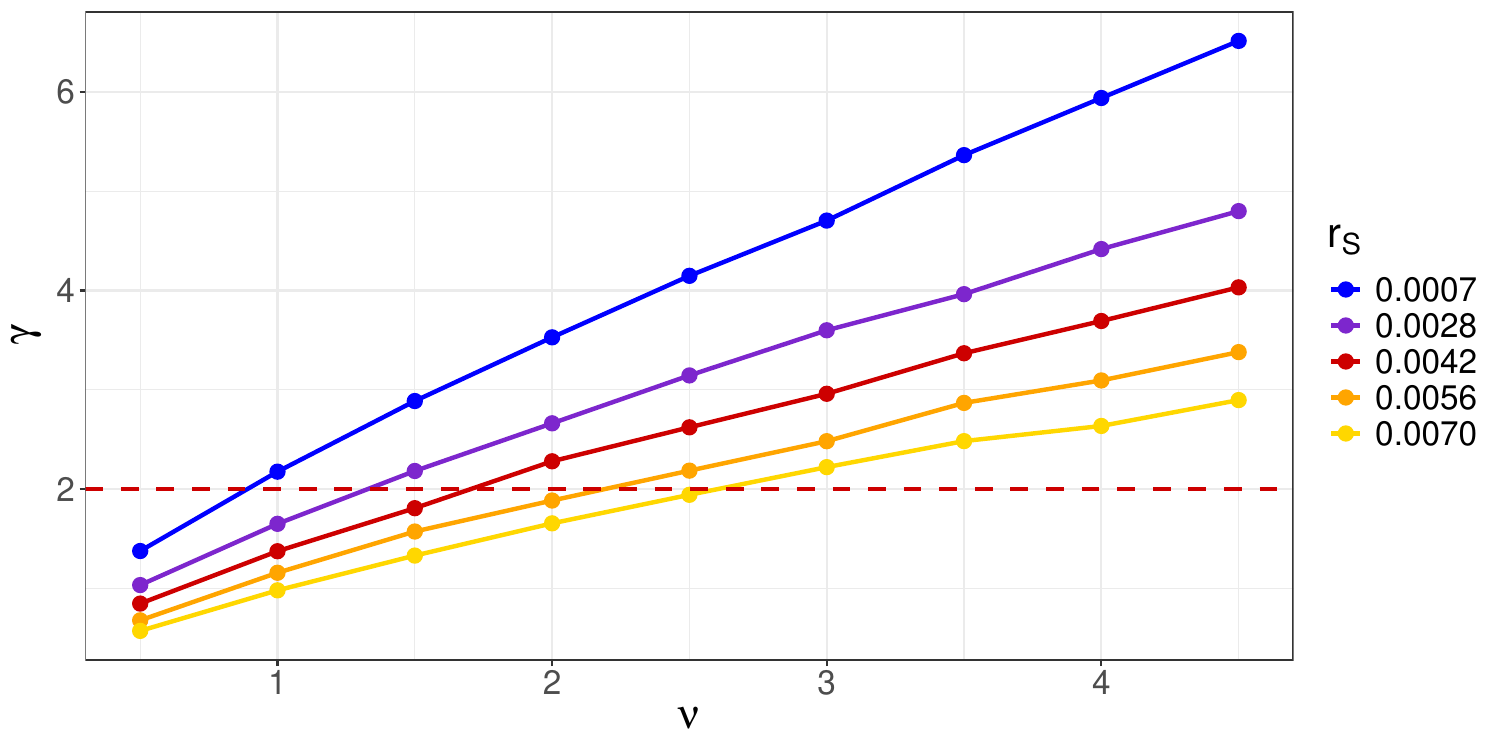}
    \caption{Sensitivity of the smoothness parameter $\gamma$. Estimates of the smoothness parameter $\gamma$ are shown for different values of the Mat\'{e}rn parameter $\nu$ and increasing separation radius $r_S$.}
    \label{fig:sepRadvsgamma}
\end{figure}

Despite their optimality, SP-based guarantees rely on smoothness and sampling conditions that can be too restrictive in practice:
\begin{enumerate}
    \item \textbf{Smoothness threshold.}
         The optimal rate in \eqref{eq:pp_rate} holds only when $\gamma > 2$. This means that sufficiently smooth processes are assumed but  ``rougher'' processes (e.g., exponential kernels with $\nu = 0.5$ and $\gamma < 2$) are excluded. Even for $\gamma = 2 + \delta$ with small $\delta > 0$, the requirement $m = O\big(n^{2/(2+\delta)}\big)$ implies a rapidly growing number of SPs, which can become computationally prohibitive for large $n$.
    \item \textbf{Design dependence and enlarging domains.}
         While the approximation $\gamma \approx 2(\nu + d/2)/d$ is appropriate under infill asymptotics, enlarging-domain designs lead to a separation radius $r_S = \frac{1}{2} \min_{i \neq j} \| s_i - s_j \|_2$ that may converge to a strictly positive constant \citep{WanJin2022}. This induces a reduction in the effective value of $\gamma$, thereby restricting the validity of SP-based theoretical guarantees. In fact, $\gamma$ may fall below $2$ even for moderate smoothness levels $\nu$ as $r_S$ grows; see Figure~\ref{fig:sepRadvsgamma} for an illustration.
\end{enumerate}

To overcome these limitations and enable scalable, accurate inference for complex spatial designs, we propose a strategy that balances statistical accuracy, computational efficiency, and robustness without requiring an unreasonable growth of $m$. To this end, we first introduce an \emph{Ensemble of Predictive Processes} (EPP) in Section~\ref{sec:EPP}, which leverages spatial partitioning to reduce computational burden while preserving accuracy, as we show in Theorem \ref{theo:conv_epp}.

\section{Ensemble of predictive processes}
\label{sec:EPP}
The EPP, formally introduced in Definition \ref{def:epp} below,  aims to overcome the limitations of PPs with SPs discussed in Section \ref{sec:limitations} by combining low-rank  approximations of GPs with a divide-and-conquer strategy through spatial partitioning. 

\begin{definition}[Ensemble of Predictive Processes]\label{def:epp}
Let $\mathcal{P} = \{\mathcal{S}_k\}_{k=1}^K$ be a partition of the spatial domain $\mathcal{S}$ into $K$, possibly overlapping, subregions. For each $\mathcal{S}_k$, let $\{\tilde{s}_{k,i}\}_{i=1}^m$ denote a set of $m$ SPs. The \mbox{EPP}($K$, $m$) at a predictive location $s_\star$ is defined as a locally weighted mixture of PPs with density  
\[
p(y(s_\star) \mid y) = \sum_{k=1}^K \pi_k(s_\star \mid \mathcal{S}_k) \, p(y(s_\star) \mid y, \mathcal{S}_k),
\]
 where $p(y(s_\star) \mid y, \mathcal{S}_k)$ is the predictive distribution of a PP on $\mathcal{S}_k$, and $\pi_k(s_\star \mid \mathcal{S}_k)$ is a local weight such that $\sum_{k=1}^K \pi_k(s_\star \mid \mathcal{S}_k)=1$. The corresponding predictive mean and variance are:
\begin{align*}
    \mu_{K,m}(s_\star) &= \sum_{k=1}^K \pi_k(s_\star \mid \mathcal{S}_k) \, \mu_m(s_\star \mid \mathcal{S}_k), \\
    \sigma^2_{K,m}(s_\star) &= \sum_{k=1}^K \pi_k(s_\star \mid \mathcal{S}_k) \Bigg[ \sigma^2_m(s_\star \mid \mathcal{S}_k) + \mu^2_m(s_\star \mid \mathcal{S}_k)\Bigg] -  \Bigg[\sum_{k=1}^K \pi_k(s_\star \mid \mathcal{S}_k)\mu_m(s_\star \mid \mathcal{S}_k)\Bigg]^2.
\end{align*}
\end{definition}
 
Section \ref{sec:bound_cont} discusses the partition of the spatial domain and guides the choice of the weights. In Section \ref{sec:theoepp}, we show that the EPP is computationally scalable while retaining theoretical convergence guarantees for the predictive mean.

\subsection{Spatial partitioning, local weights, and continuity}
\label{sec:bound_cont}
The choice of partition $\mathcal{P}$ strongly affects both computational efficiency and predictive accuracy. Simple schemes, such as regular grids, generally lead to imbalanced or empty regions. We instead adopt a principled approach based on \emph{Support Points-based Voronoi Tessellation} (SPVT), which combines SP optimality with Voronoi partitioning \citep{DuFabGun1999}. SPVT proceeds  as follows (see also Algorithm D.1 in the Supplementary Material for details):
\begin{enumerate}
    \item \textbf{Site selection}. Choose $K$ sites $\{u_k\}_{k=1}^K$ as SPs over $\mathcal{S}$.
    \item \textbf{Voronoi regions}. Each site $u_k$ induces a Voronoi region (or cell) $\mathcal{V}_k$, for $k=1,\ldots,K$:
    \begin{equation*}
        \mathcal{V}_k=\{s\in\mathcal{S}\mid \Vert s-u_k\Vert_2 < \Vert s-u_j\Vert_2, \forall j\neq k\}.
    \end{equation*}
    \item \textbf{Introduce overlap}. For $\delta>0$, define overlapping regions
    \begin{equation*}
        \mathcal{S}_k=\{s\in\mathcal{S}\setminus\mathcal{V}_k\mid \min_{x\in\mathcal{V}_k}\Vert s-x\Vert_2 \leq \delta\}\cup\mathcal{V}_k.
    \end{equation*}
    \item \textbf{Local SP selection}. Within each  $\mathcal{S}_k$, select $m$ SPs to construct the local PP.
\end{enumerate}

SPVT yields balanced, density-aware partitions \cite[cf.,][]{Sec2002,BalHec2008} and satisfies Assumption \ref{ass:partitioning}(B) required for our theoretical results (Section~\ref{sec:theoepp}). In particular, SPVT ensures $|\mathcal{S}_k|\approx[K\,f(u_k)]^{-1}$, so that $\mathbb{P}_F(s \in \mathcal{S}_k) \approx 1/K$. Figure~C.1 in the Supplementary Material provides an illustration of SPVT for several spatial designs $F$. 

In Definition \ref{def:epp}, a natural choice for disjoint partitions is  $\pi_k(s_\star\mid\mathcal{S}_k)=\mathbb{I}(s_\star\in\mathcal{S}_k)$, which has favorable asymptotic properties \citep{SzaHadVan2025} but induces discontinuities at region boundaries in finite samples. Allowing overlaps helps achieve continuity, and we adopt soft weights based on truncated localization kernels \citep{HorBar2025},
\[
\pi_k(s_\star\mid\mathcal{S}_k) \propto K(s_\star,u_k)\,\mathbb{I}(s_\star\in\mathcal{S}_k),
\]
where $u_k$ is the centroid of $\mathcal{S}_k$ and $K$ is a distance-decaying kernel. Let $\Delta(\mathcal{S}_k)$ denote the boundary of region $k$. To ensure continuity, $K$ must satisfy, for any $s_0\in\Delta(\mathcal{S}_k)$,
\[
\lim_{\substack{s_\star\to s_0 \\ s_\star\in \mathcal{S}_k}} K(s_\star,u_k)=0.
\]
A Gaussian kernel with mean $u_k$ and variance tied to the distance to the boundary,
\[
K(s_\star,u_k)=\exp\!\left\{-\frac{\|s_\star-u_k\|_2^2}
{\operatorname{dist}(s_\star,\Delta(\mathcal{S}_k))}\right\},
\qquad
\operatorname{dist}(s_\star,\Delta(\mathcal{S}_k))
= \min_{s_0\in\Delta(\mathcal{S}_k)}\|s_\star-s_0\|_2^2,
\]
satisfies this condition. Figures~C.3–C.4 in the Supplementary Material illustrate the effect of different kernels on the ensemble mean and local weights.

\subsection{Theoretical properties}\label{sec:theoepp}
Theorem \ref{theo:conv_epp} extends  the convergence rate of $\mu_m$ of a PP to the predictive mean $\mu_{K,m}$ of an EPP$(K,m)$.  Let $\alpha\in[0,1)$ determine the number of partitions $K=O(n^\alpha)$ and the number of SPs per region $m=O(n^{(1-\alpha)2/\gamma})$. We assume Assumption~\ref{ass:regularity} and  the additional conditions below.

\begin{assumption}\label{ass:partitioning}
Let the following hold:
\begin{enumerate}[(A)]
    \item The spatial domain satisfies $\mathcal{S} = \bigcup_{k=1}^K \mathcal{S}_k$, where overlaps are allowed.
    \item The probability $\mathbb{P}_F(s \in \mathcal{S}_k) = O(1/K)$.
\end{enumerate}
\end{assumption}

\begin{theorem}\label{theo:conv_epp}
Under Assumptions \ref{ass:regularity}--\ref{ass:partitioning}, if $\gamma>2$,  the predictive mean $\mu_{K,m}$ of the EPP satisfies:
\[
\| \mu_{K,m} - \mu_0 \|_{L^2(F)}^2 = O_p\left(n^{-(1-\alpha)\left(1 - \frac{1}{\gamma}\right)}\right).
\]
\end{theorem}
The proof of Theorem \ref{theo:conv_epp} is provided in  Supplementary Material B.1. 

Compared to the PP rate in \eqref{eq:pp_rate}, the EPP introduces $\alpha$ as a tunable parameter that explicitly controls the trade-off between scalability ($\alpha\to1$) and accuracy ($\alpha\to0$). Increasing $\alpha$ yields more partitions and smaller regions, enabling parallelization but slowing convergence. For $\alpha=0$, EPP reduces to a single PP.

\paragraph{Theoretical and practical implications.}
Computing an EPP requires evaluating $K=O(n^\alpha)$ local PPs, each based on $m=O(n^{(1-\alpha)2/\gamma})$ SPs. This alleviates the issue that, for PPs, achieving convergence when $\gamma=2+\delta$ and small $\delta>0$ requires the number of SPs to grow almost linearly with $n$. By contrast, EPP remains computationally feasible while still ensuring convergence through an appropriate choice of $\alpha$, and thus avoids the near-linear growth in $m$ required by global PPs. In addition, computations can be parallelized, enabling substantial speed-ups in practice. This computational advantage becomes particularly evident in high-dimensional settings, as demonstrated in Section~\ref{sec:num}. Finally, when $1<\gamma\le2$, asymptotic convergence cannot be guaranteed. However, the following Corollary helps to explain  the empirical improvements observed in our experiments in Section~\ref{sec:num} in cases with enlarging domain and $1<\gamma\le2$.

\begin{corollary}\label{theo:conv_epp_gamma}
Under Assumptions \ref{ass:regularity}--\ref{ass:partitioning}, if $K = n/m$, $m$ fixed,  and $\{\tilde{s}_{k,i}\}_{i=1}^m = \{s_i\in\mathcal{S}: s_i\in\mathcal{S}_k\}$ it holds that $\| \mu_{K,m} - \mu_0 \|_{L^2(F)}^2=O(1)$ and it is upper-bounded by
\[
\| \mu_{K,m} - \mu_0 \|_{L^2(F)}^2 \leq c_1 m^{-(1-1/\gamma)},
\]
for some constant $c_1>0$ and $\gamma>1$.
\end{corollary}

The proof of Corollary \ref{theo:conv_epp_gamma} is provided in Supplementary Material B.2. In contrast to PPs, whose theory requires $\gamma>2$ (see Section \ref{sec:limitations}), the predictive error of the EPP can still be characterized when $\gamma\le2$ by considering the extreme case $\alpha=1$. 

Table \ref{tab:rates} contrasts key theoretical properties of the GP, PP and EPP. 
\begin{table}[!ht]
  \centering
  \resizebox{1\textwidth}{!}{
    \renewcommand{\arraystretch}{1.35}
    \begin{tabular}{ccccccccccc}
    \toprule
         & & GP & & PP with SPs \citep{SonDaiGen2025} & & EPP (Theorem \ref{theo:conv_epp}) & & EPP (Corollary \ref{theo:conv_epp_gamma}) \\
        \cmidrule{3-3}\cmidrule{5-5}\cmidrule{7-7}\cmidrule{9-9}
         Smoothness & & $\gamma > 1$ & & $\gamma > 2$ & & $\gamma > 2$ & & $1 < \gamma \leq 2$ \\
         Partitions &  & --- & &--- & &$K=O(n^\alpha)$, $\alpha\in[0,1)$ & & $K=n/m$\\
         Support points & & --- & & $m=O(n^{2/\gamma})$ &  & $m = O(n^{(1-\alpha)2/\gamma})$, $\alpha\in[0,1)$ & & $m$ fixed \\
         Convergence rate & & $O_p(n^{-(1 - 1/\gamma)})$ & & $O_p(n^{-(1 - 1/\gamma)})$ &  & $O_p(n^{-(1-\alpha)(1 - 1/\gamma)})$ & & $O_p(1)$ \\
         Computations & & $O(n^3)$ & & $O(n^{1 + 4/\gamma})$ & & $O(n^{1 + (1-\alpha)4/\gamma})$ & & $O(n)$ \\
    \bottomrule
    \end{tabular}%
  }
  \vspace{0.25em}
\caption{Summary of  key theoretical properties of the GP, PP with SPs, and EPP. The smoothness conditions under which the theoretical results hold are reported in the first row. The second and third row highlight the order of  partition size and number of SPs in the respective prediction errors. Given these, the last two rows summarize the convergence rates and computational costs.}\label{tab:rates}%
\end{table}%

\section{Multi-resolution ensemble of predictive processes}
\label{sec:MREPP}
The theoretical results in Section~\ref{sec:theoepp} and \citet{SonDaiGen2025} show that an accurate approximation of the full GP requires the number of inducing points $m$ to grow with $n$. Consequently, projecting onto a fixed low-rank subspace limits the ability to capture fine-scale dependence and typically results in oversmoothing, particularly at locations far from inducing points. 
We turn this limitation into an advantage. In Section \ref{subsec:robustnessPPEPP}, we show that PPs and EPPs with fixed and small $m$ reduce the generalization error under contaminated training data in a finite-sample regime. In practice, however, the generalization error can only be upper bounded, and the optimal value of $m$ cannot be computed in practice. This motivates the definition of a multi-resolution ensemble of PPs and EPPs in Section \ref{subsec:MREPP} that mitigates these challenges for finite $n$. 

\subsection{Predictive robustness of PPs and EPPs}\label{subsec:robustnessPPEPP}
After the formal definition of training data contamination,  we study the propagation of the influence of contamination on the predictive mean, and derive an upper bound for the respective generalization error. In what follows, the results are stated for the more general EPP($K$,$m$) since those for the single PP can be recovered as a special case setting $K=1$. As before, we denote by $\mu_{K,m}(s_\star;y)$ the predictive mean of the EPP($K$,$m$) at a new location $s_\star$ trained on data $y$. In line with the motivation of the multi-resolution approach in Section \ref{sec:MREPP}, we assume $n$ to be finite. 

\begin{definition}[Training Data Contamination]\label{def:datacon}
   Let $\Delta=(\Delta_1,\ldots,\Delta_n)^\top\in\mathbb{R}_{<\infty}^n$ be  the \emph{contamination vector} and $y^\Delta=y+\Delta$ the  \emph{contaminated data}. We then call  $\mu_{K,m}^\Delta(s_\star;y+\Delta)=\mu_{K,m}(s_\star;y)+\mu_{K,m}(s_\star;\Delta)$ \emph{prediction under data contamination} and  the term $\mu_{K,m}(s_\star;\Delta)$ \emph{contamination effect}. Due to the additive decomposition of the training data under contamination and its prediction, the respective generalization error  $\| \mu_{K,m}^\Delta - \mu_0 \|_{L^2(F)}^2$ is decomposed as
  \begin{align}\nonumber
    \| \mu_{K,m}^\Delta - \mu_0 \|_{L^2(F)}^2 &= \underbrace{\int_\mathcal{S}\left[\mu_{K,m}(s;y)-\mu_0(s)\right]^2 dF(s)}_{\text{Approximation error }A_{K,m}} + \;2\underbrace{\int_\mathcal{S}\left[\mu_{K,m}(s;y)-\mu_0(s)\right]\mu_{K,m}(s;\Delta) dF(s)}_{\text{Bias-Contamination interaction }B_{K,m}}\\\label{eq:GEcon}
    &\qquad + \underbrace{\int_\mathcal{S}\left[\mu_{K,m}(s;\Delta)\right]^2 dF(s).}_{\text{Integrated squared contamination effect }C^2_{K,m}(\Delta)}
\end{align} 
\end{definition}
The generalization error in Definition \ref{def:datacon} does not have a closed-form  solution. However, we can provide an upper bound for $\| \mu_{K,m}^\Delta - \mu_0 \|_{L^2(F)}^2$ by studying the three terms in \eqref{eq:GEcon} in more detail.  
The first term $A_{K,m}$ is the approximation error induced by the EPP($K$,$m$). It is decreasing in $m$ and  minimized when using a full GP in each region, that is, by choosing $m=n/K$.
The bias-contamination interaction $B_{K,m}$ can be bounded in absolute terms by   $|B_{K,m}|\leq A_{K,m}^{1/2}(C^2_{K,m}(\Delta))^{1/2}$. Hence, for upper-bounding \eqref{eq:GEcon} it remains to study the behavior of the integrated squared contamination effect $C^2_{K,m}(\Delta)$. In Section B.3 in the Supplementary Material, we show that $C^2_{K,m}(\Delta) \lesssim \min(m, m_0)/n\Vert\Delta\Vert_2^2$ is non-decreasing in $m$, where $m_0=(K/(n\tau^2))^{-1/\gamma}$. Hence, minimizing the generalization error $ \| \mu_{K,m}^\Delta - \mu_0 \|_{L^2(F)}^2$ can be seen as a trade-off between approximation accuracy ($m\to  n/K$) and predictive robustness ($m < n/K$).

In Proposition \ref{prop:upperbound} we derive an $m$-dependent upper bound $\eps^+(m)$ on  \eqref{eq:GEcon} and characterize the optimal choice of  $m^\ast_+$, that minimizes $\eps^+(m)$.

\begin{proposition}\label{prop:upperbound}
    Assume Assumptions \ref{ass:regularity} and \ref{ass:partitioning} hold. Then, for large, but finite sample size $n\geq n_0$, and a positive constant $c_\eps>0$ it holds that 
    $\| \mu_{K,m}^\Delta - \mu_0 \|_{L^2(F)}^2 \lesssim \eps^+(m)$, where
    \begin{align*}
    \eps^+(m) = c_\eps\left(\frac{m^{(1-\gamma)/2}}{\sqrt{\gamma-1}}+\frac{\min(\sqrt{m},\sqrt{m_0})\Vert\Delta\Vert_2}{\sqrt{n}}\right)^2,\qquad m_0=\left(\frac{K}{n\tau^2}\right)^{-1/\gamma}.
    \end{align*}
    Moreover, $\eps^+(m)$ is uniquely  minimized at
    \[
    m_+^\star =
    \begin{cases}
    (\gamma - 1)^{1/\gamma}\left(\dfrac{n}{\|\Delta\|_2^2}\right)^{1/\gamma}, & 0 < m_+^\star < m_0, \\
    n/K, & m_0 \leq m_+^\star \leq n/K.
    \end{cases}
\]
\end{proposition}
Proposition \ref{prop:upperbound} highlights the dependence of $\eps^+(m)$ and the optimal $m^\ast_+$ on the signal-to-contamination ratio $n/\|\Delta\|_2^2$, the number of regions $K$, and the smoothness of the process $\gamma$:
\begin{itemize}
    \item $m^\ast_+$ is proportional to $n/\|\Delta\|_2^2$. Higher training data contamination compared to the sample size requires fewer SPs to limit the effect on predictions.
    \item Since we assume $\Delta$ to be finite, it holds that  $n/\|\Delta\|_2^2\to \infty$ and $m^\ast_+\to\infty$ for $n\to\infty$. Hence, to study predictive robustness we focus on the case where $n$ is large but finite $ n_0<n<\infty$. 
    \item If $K\to  n$ then $m^\ast_+\to 1$ (each region contains fewer and fewer observations). Consequently, the more partitions we assume, the less robust the EPP can be.
    \item If $\gamma\to \infty$ then  $m_0\to  1$ and hence $m^\ast_+=n/K$. Consequently, the smoother the process, the less \eqref{eq:GEcon} is affected by training data contamination.
\end{itemize}
In practice, since the true smoothness and the level of training data contamination are unknown, the optimal $m_+^\ast$  that minimizes the upper bound $\eps^+(m)$ cannot be computed. However, we can control the number of regions $K$. This motivates the definition of an ensemble of EPPs over multiple  pairs $\{(K_l,m_l)\}_{l=1}^L$, with an increasing number of ensembles $K_l$ and a decreasing number of SPs $m_l$, which we propose in the next section.  

\subsection{Multi-resolution ensemble of PPs}\label{subsec:MREPP}
In this section, we introduce the \emph{Multi-Resolution Ensemble of Predictive Processes} (MREPP) that avoids the need to explicitly compute the optimal $m$, while  balancing approximation accuracy and predictive robustness. MREPP constructs ensembles across multiple spatial resolutions, ranging from possibly overlapping coarse to fine partitions, and aggregates their predictions. We provide a formal definition next. 

\begin{definition}[Multi-Resolution Ensemble of Predictive Processes]\label{def:MREPP}
Let $\mathcal{P}_l = \{\mathcal{S}_{l,k}\}_{k=1}^{K_l}$ be a (possibly overlapping) partition of  $\mathcal{S}$ at \emph{resolution}  $l = 1,\ldots,L$,  such that $K_1 < \cdots < K_L$. For each  $\mathcal{S}_{l,k}$, let $\{\tilde s_{l,k,i}\}_{i=1}^{m_l}$ be a set of $m_l$ SPs such that $m_1 \geq \ldots \geq m_L$. Then, the MREPP($L$,$\{K_l\}_{l=1}^L$,$\{m_l\}_{l=1}^L$) of resolution depth $L$ is defined as
\[
p(y(s_\star) \mid y) = \sum_{l=1}^L \sum_{k=1}^{K_l} \pi_{l,k}(s_\star \mid \mathcal{S}_{l,k}) \, p(y(s_\star) \mid y, \mathcal{S}_{l,k}),
\]
where $\pi_{l,k}(s_\star \mid \mathcal{S}_{l,k})$ are resolution-specific local weights satisfying $\sum_{l=1}^L \sum_{k=1}^{K_l} \pi_{l,k}(s_\star \mid \mathcal{S}_{l,k}) = 1$. The predictive mean and variance are
\begin{align*}
    \mu_{L,K,m}(s_\star) &= \sum_{l=1}^L \sum_{k=1}^{K_l} \pi_{l,k}(s_\star \mid \mathcal{S}_{l,k}) \, \mu_m(s_\star \mid \mathcal{S}_{l,k}), \\
    \sigma^2_{L,K,m}(s_\star) &= \sum_{l=1}^L \sum_{k=1}^{K_l} \pi_{l,k}(s_\star \mid \mathcal{S}_{l,k}) \Bigg[ \sigma^2_m(s_\star \mid \mathcal{S}_{l,k}) + \mu^2_m(s_\star \mid \mathcal{S}_{l,k})\Bigg] \\
    &\qquad-  \Bigg[\sum_{l=1}^L \sum_{k=1}^{K_l} \pi_{l,k}(s_\star \mid \mathcal{S}_{l,k})\mu_m(s_\star \mid \mathcal{S}_{l,k})\Bigg]^2.
\end{align*}
\end{definition}

The spatial weights  $\pi_{l,k}(s_\star\mid \mathcal{S}_{l,k})$ are defined as the normalized  products of horizontal and vertical weights $\pi_{l,k}(s_\star\mid \mathcal{S}_{l,k})\propto p(l)\pi_k(s_\star\mid \mathcal{S}_{l,k})$. The horizontal weights $\pi_k(s_\star\mid \mathcal{S}_{l,k})$ use truncated localization kernels from Section~\ref{sec:bound_cont}. The vertical weights $p(l)$, collected in the vector $\hat{p}$, are learned by minimizing an empirical version of the generalization error on a calibration set:
\begin{equation}\label{eq:pl}
    \hat{p} = \arg\min_{p\in\mathcal{P}} \frac{1}{n_2}\sum_{i=1}^{n_2} \big(\mu_{L,K,m}(s_{2,i}|y_1),\,y_2(s_{2,i})\big)^2,\qquad \mathcal{P}=\left\{p\in(\mathbb{R}^+)^L: \|p\|_1=1\right\},
\end{equation}
where $y_1=\lbrace y(s_{1,i})\rbrace_{i=1}^{n_1}$ and $y_2=\lbrace y(s_{2,i})\rbrace_{i=1}^{n_2}$, with $n_1+n_2=n$, denote a disjoint division of the observed data into training and calibration subsets, respectively. In this way, the vertical weights adaptively govern the contribution of each resolution: The sequence of coarse to fine partitions progressively transitions from PPs that capture global spatial trends to those that capture increasingly local features, thereby moving from more robust but potentially oversmoothed predictions, to more accurate but potentially overfitted ones. The idea is that if contaminated data are present in the training set, coarser resolutions should be upweighted to avoid overfitting in those regions.  Algorithm D.2 in the Supplementary Material summarizes spatial prediction with the MREPP. 

The MREPP recovers the PP when $L=1,K_1=1$ and the EPP when $L=1,K_1>1$. It is computationally efficient because computations are parallelizable across resolutions and subregions. Its multi-resolution design enables predictive robustness under training data contamination, while preserving local accuracy if contamination is absent, as we formalize in the following proposition. 

\begin{proposition}\label{prop:mrepp-opt}
    Assume  Assumptions \ref{ass:regularity} hold. Consider the MREPP($L$,$\{K_l\}_{l=1}^L$,$\{m_l\}_{l=1}^L$) with predictive mean $\mu_{L,K,m}(s)$ where the resolution weights are defined as in \eqref{eq:pl}, and $s_{2,i}\sim F(s)$. Then, for a calibration set large enough $n_2\geq n_0$:
    \[
    \| \mu_{L,K,m}^\Delta - \mu_0 \|_{L^2(F)}^2 \leq \min_{l\in\{1,\ldots,L\}} \| \mu_{K_l,m_l}^\Delta - \mu_0 \|_{L^2(F)}^2 + O_p(n_2^{-1/2}).
    \]
\end{proposition}
Proposition \ref{prop:mrepp-opt} shows that the MREPP has  predictive performance that is at least as good as that of any EPP$(K_l,m_l)$, $l=1,\ldots,L$, since the generalization error of the MREPP is bounded from above by the generalization error of the best performing EPP($K_l,m_l$). 

\subsection{Tuning MREPP}\label{sec:tune-MREPP} 
Although theoretical results can guide the optimal choice of the number of SPs in each subregion, choosing $L$, and the sequences $K_1<\ldots <K_L$, and $m_{\max}\geq m_1\geq\ldots\geq m_L$ remain a practical question specific to the data and the problem to be solved. We therefore empirically investigate   the sensitivity of the MREPP to these choices in the Supplementary Material E, with results summarized in Figure E.4. The main findings from these experiments are as follows. 

First, increasing the number of resolutions $L$ consistently improves predictive performance and robustness. Among the conducted experiments with varying sample sizes and contamination levels, larger $L$ performed best and was the most stable. These benefits of larger $L$ become even more evident in the real data applications in Section~\ref{sec:appl}, in particular for large datasets (see Table \ref{tab:resappl}). These results also highlight  that restricting attention to only the coarsest resolution (i.e., the PP) and the finest one is suboptimal in practice, especially for density prediction. Second,  smaller $m_{\max}$ performs better for small sample sizes $n$, while larger $m_{\max}$ offers improvements for larger datasets. In line with our theoretical results, these results indicate that $m_{\max}$ should be much smaller than $n$ to improve predictive robustness. For example, for $n=1000$  we find $m_{\max} = 50$ outperforms $m_{\max} = 200$ empirically.
Finally, as expected, increasing $L$ means higher runtime, while runtime  decreases with smaller $m_{\max}$. 

Taking all aspects into account, as a practical recommendation, we suggest using large $L$ for stability, and choosing $m_{\max}$ small relative to $n$ to preserve predictive robustness, while conserving computational resources.

\section{Empirical evaluation}
\label{sec:num}
In this section, we conduct experiments to evaluate the proposed MREPP and EPP in terms of predictive accuracy, uncertainty quantification, robustness to data contamination, and computational efficiency on synthetic data in Section \ref{sec:sim} and on real data in Section \ref{sec:appl}. The computations were implemented in \textsf{R}~4.4.0 and executed on a compute node equipped with dual Intel\textsuperscript{\textregistered} Xeon\textsuperscript{\textregistered} Platinum and 512~GB RAM.

\subsection{Simulation study}
\label{sec:sim}

\paragraph{Data generation.} Data are generated from  \eqref{eq:gp} with locations  drawn uniformly over $\mathcal{S}$ in three scenarios. Scenarios 1 and 2 assume fixed space $\mathcal{S}=[-3,3]^2$ and  enlarging space with $n$ such that the separation radius $r_S=0.001$ is constant, respectively. Scenario 3 considers fixed space $\mathcal{S}=[-3,3]^2$ and in order to mimic data contamination,  we sequentially set $1\%$ of the observations to $5,10,15$, which correspond to $\{10,20,30\}$ times the true standard deviation $\sqrt{0.25}$. 

In each scenario, the true GP assumes a Mat\'{e}rn kernel with parameters $(\phi,\nu)=(0.21,1.5)$, marginal variance $\eta^2=1.5$, and Gaussian nugget $\varepsilon(s)\sim N(0,0.25)$. In Scenarios 1 and 2 we additionally consider $(\phi,\nu)=(0.33,0.5)$ to contrast performance with a less smooth process. The range parameter $\phi$ is chosen such that the effective range of the GP is equal to 1. For all scenarios, we consider $n=\{1000,2500,5000,7500,10000\}$ locations for training and an additional test set of $N=1000$ observations for prediction. We generate $R=100$ data replications, where for each replicate, the  sequences $\{(s_i, w(s_i), \varepsilon(s_i))\}_{i=1}^n$ and $\{(s_{\star,i}, w(s_{\star,i}), \varepsilon(s_{\star,i}))\}_{i=1}^N$ for the training and test sets, respectively, are resampled.

\paragraph{Benchmark methods.}
As competitors, we include the full GP, the PP of \citet{SonDaiGen2025}, and the local approximate GP (LAGP) of \citet{GraApl2015}. The PP is a global approximation method, whereas LAGP adopts a local approximation strategy, representing fundamentally different approaches to scalable spatial prediction. Together, they cover a broad spectrum of GP approximation strategies, and both have been shown to perform well in the spatial prediction literature, thereby constituting a well-established set of competitors. All methods assume a Mat\'ern kernel with parameters fixed at their true values to isolate methodological differences. We come back to this point in Section \ref{sec:concl}.

\paragraph{Hyperparameter settings.} 
For the EPP, we consider three values of $\alpha \in \{0.2, 0.4, 0.5\}$, set $K = n^{\alpha}$, and determine the number of SPs using the theoretical recommendation when $\gamma > 2$ (i.e., $\nu=1.5$), while for $\gamma \leq 2$ (i.e., $\nu=0.5$) we instead use a fraction of the expected number of observations in the corresponding region. Specifically, we choose $m = \min\{ (n/K)^{2/\gamma},\ 0.5\,(n/K) \}$.

For  MREPP, we investigate three settings with different $L$ and varying $\alpha=(\alpha_1,\ldots,\alpha_L)$ (and therefore $K_l = n^{\alpha_l}$), where $\alpha_l$ is the value used for the $l$-th EPP in the MREPP construction. Specifically, we  consider $L = 2$ and $\{\alpha_l\}_{l=1}^2 \in \{0, 0.5\}$, $L = 4$ and $\{\alpha_l\}_{l=1}^4 \in \{0, 0.2, 0.4, 0.5\}$, and $L = 6$ and $\{\alpha_l\}_{l=1}^6 \in \{0, 0.2, 0.3, 0.4, 0.45, 0.5\}$. We discuss the tuning of MREPP in Section \ref{sec:tune-MREPP}. The number of SPs for each $l=1,\ldots,L$ in the MREPP follows the same principle used for the EPP, but we impose an upper bound on the number of SPs, as discussed in Section \ref{sec:tune-MREPP}, $m_l = \min\{ m_{\max},\ (n/K_l)^{2/\gamma},\ 0.5\,(n/K_l) \}$. Here we choose $m_{\max}=200$. To obtain data-driven estimates of the resolution weights $p(l)$, we employ a calibration set of size $0.2 n$. This yields weight estimates that reflect local predictive performance while maintaining computational efficiency.

Both EPP and MREPP allow for a small overlap between adjacent regions to ensure continuity at the boundaries. Horizontal weights are defined using the distance-dependent truncated Gaussian localization kernel introduced in Section~\ref{sec:bound_cont}.
Throughout, we refer to EPP as EPP$(\alpha)$ and  MREPP as MREPP$(L)$, respectively.

The PP is estimated using $m = \min\{ n^{2/\gamma},\ 0.5 n \}$.
The corresponding values of $\gamma$ and $m$ are provided in Table~E.1 of the Supplementary Material.

We implement LAGP using the active learning Cohn (ALC) strategy \citep{Coh1996} to construct the local   prediction design. For each predictive location, ALC identifies a local subset of points that minimizes the predictive variance. Following \citet{GraApl2015}, we initialize the search with the five nearest neighbors and set $k_{\max} = 50$.

For $n>5000$ we found it computationally infeasible to run the full GP with all $R$ replicates. We therefore refrain from reporting performance metrics for the full GP and large sample sizes except for the runtime which we report with 10 replicates for the full GP. Furthermore, although the MREPP, EPP, and LAGP procedures are in principle parallelizable, we execute all methods in serial to facilitate a fair comparison with the full GP and PP.

\paragraph{Evaluation metrics.}
To assess the performance of each method, we consider the following metrics. As a measure of point prediction accuracy, we compute the predictive Root Mean Squared Error (RMSE) on the test set. We complement it with the Log Predictive Score (LPS) to quantify probabilistic prediction accuracy. Then, we investigate empirical coverage and average width of 90\% highest posterior density (HPD) intervals in the training data, as well as the predictive interval score \citep{GneRaf2007} on the test set to balance coverage and sharpness. Finally, we report the total runtime in seconds.

\paragraph{Results.}

We focus on results  for $n=1000,5000,10000$ in the following and present  further results for all $n$ in Supplementary Material E.

\begin{figure}[t]
    \centering
    \includegraphics[width=1\linewidth]{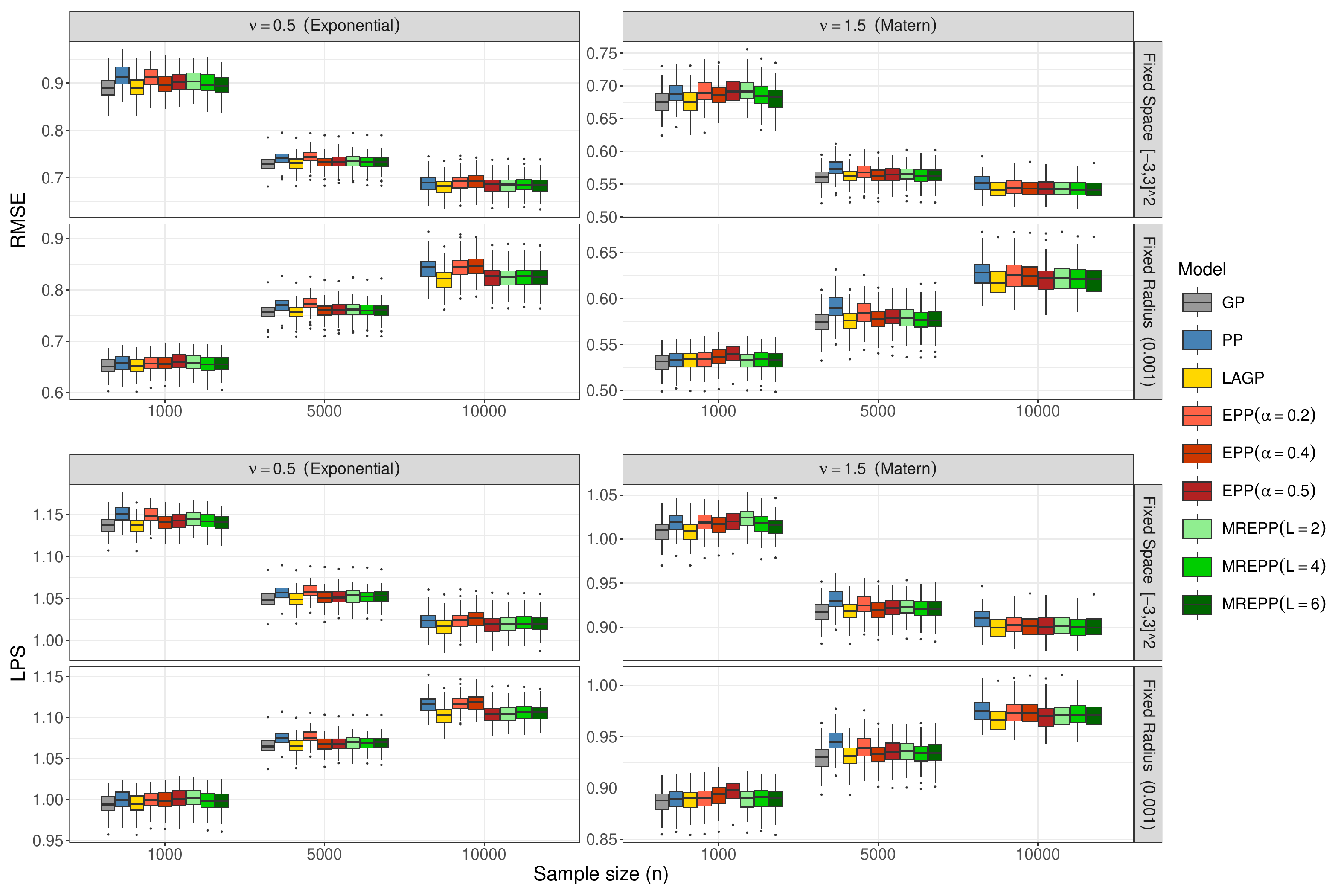}
    \caption{Scenarios 1 and 2: Point and probabilistic predictions. The RMSE (top panels) and the LPS (bottom panels) are shown for fixed space (Scenario 1) and enlarging domain with fixed separation radius (Scenario 2) and increasing sample sizes $n=1000,5000,10000$. For both metrics, a smaller value indicates better performance.}
    \label{fig:pred-correct}
\end{figure}

\noindent\emph{Prediction accuracy and robustness}.
 Figure \ref{fig:pred-correct} shows that the  MREPP yields comparable performance to full GP (when available) and LAGP in terms of both point prediction and probabilistic prediction, for all three choices of $L=2,4,6$ in Scenarios 1 and 2.  For EPP,  the performance for $\alpha=0.5$ and $\gamma>2$ is satisfactory across all scenarios as somewhat expected. However, for $\gamma < 2$ the performance drops rapidly the smaller $\alpha$ gets. A similar behavior is observed for PP.

\begin{figure}[t]
    \centering
    \includegraphics[width=1\linewidth]{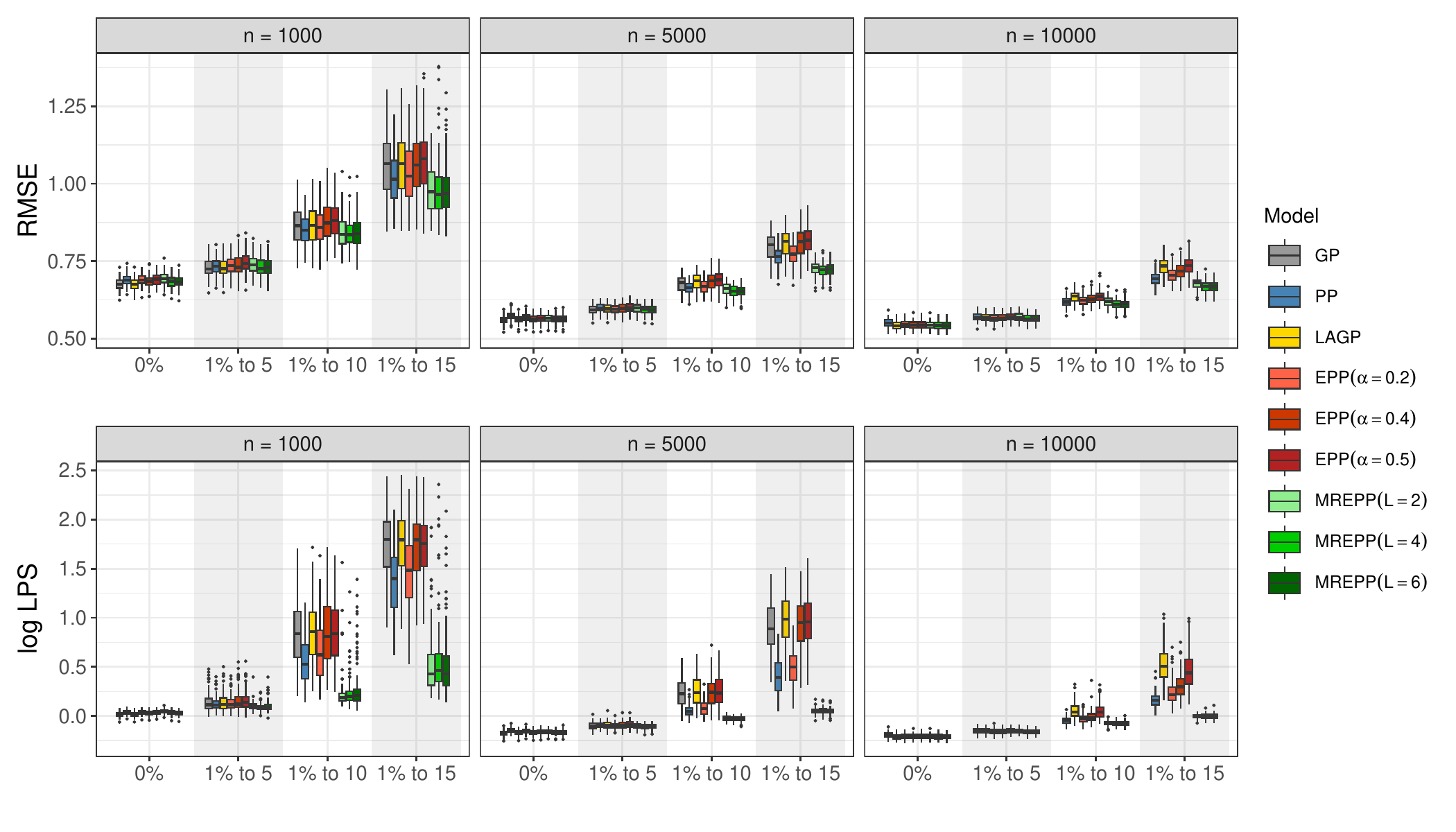}
    \caption{Scenario 3: Point and probabilistic predictions. The RMSE (top panels) and logarithm of the LPS (bottom panels) are shown for increasing level of contamination ($x$-axis) and sample sizes $n\in\{5000,7500,10000\}$.}
    \label{fig:pred-cont}
\end{figure}

\begin{figure}[t]
    \centering
    \includegraphics[width=1\linewidth]{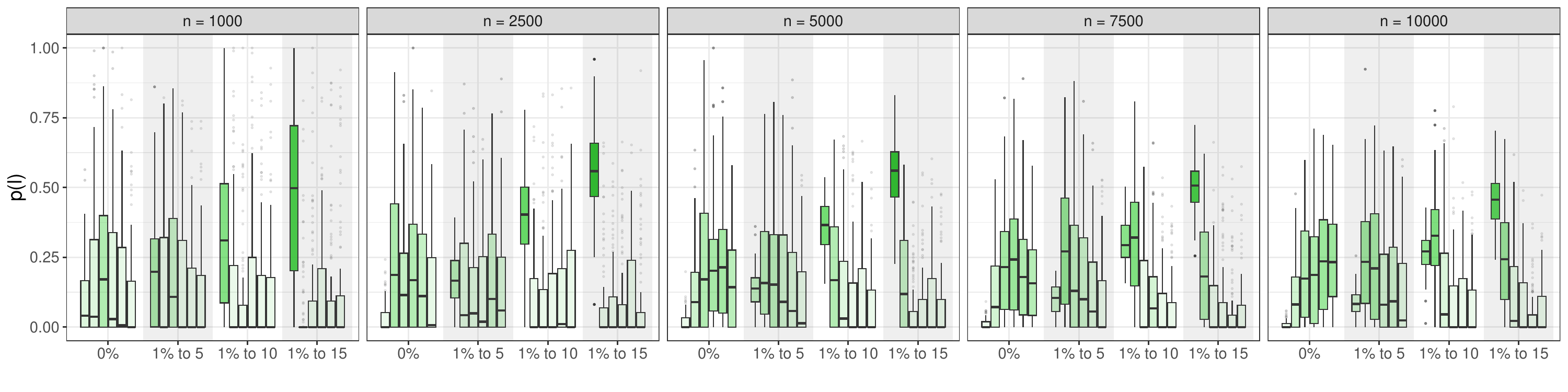}
    \caption{Scenario 3: Estimated resolution weights for MREPP($L=6$). The estimated distribution of weights $p(l)$ in MREPP is shown for increasing level of contamination and sample sizes. Each boxplot is the distribution of $p(l)$ over the replicates for $l=1,\ldots,L$ and for a given level of contamination they are sorted from left $l=1$ (coarsest resolution) to right $l=L$ (finest resolution).}
    \label{fig:pL-cont}
\end{figure}

Figure \ref{fig:pred-cont} shows that MREPP performs better than full GP, LAGP, and EPP  in terms of both point and probabilistic prediction in the presence of training data contamination as represented by Scenario 3.  The PP and EPP with smaller $\alpha$ also show satisfactory performance. However, the MREPP comes with the advantage of choosing more than one resolution, and this can be effective in achieving predictive robustness. This is indicated by Figure \ref{fig:pred-cont}  showing  that larger $L$ provides better RMSE, particularly for larger $n$.   Figure~\ref{fig:pL-cont} reports how the resolution weights change as the level of contamination increases for $L=6$ (the corresponding plots for $L=2,4$ are provided in Figure~E.3 of the Supplementary Material). When no contamination is present, MREPP allocates most weight to the finer resolutions, enabling it to capture local structure effectively. As contamination increases, however, the weights shift toward coarser resolutions, thereby reducing the influence of local outliers and improving recovery of global trends.

\noindent\emph{Uncertainty quantification}.
In Scenarios 1 and 2, the coverage of the predictive $90\%$ HPD intervals is reported in Table E.2 and is, on average, satisfactory for all methods. This indicates that, in general, the procedures are capable of producing predictive intervals with appropriate frequentist calibration. Figure E.1 presents a graphical summary of the interval scores, simultaneously assessing calibration and sharpness, and indicates that the methods perform comparably well.

However, Table \ref{tab:cov-cont} (and Figure E.2 in the Supplementary Material) highlight the superior performance of MREPP relative to the competing approaches in terms of predictive uncertainty in Scenario 3. In particular, as the level of contamination increases, the coverage of the predictive $90\%$ HPD  intervals deteriorates more slowly for the MREPP than for all benchmark methods. This behavior suggests superior robustness of  the MREPP to contamination effects. Notably, this advantage persists for all values of $L$.

\begin{table}[!ht]
  \centering
  \resizebox{1\textwidth}{!}{
    \begin{tabular}{rrrrrrrrrrrrrrrr}
        \toprule
          &       & \multicolumn{4}{c}{GP}        &       & \multicolumn{4}{c}{PP}        &       & \multicolumn{4}{c}{LAGP} \\
    \multicolumn{1}{l}{Contamination} &       & $0\%$     & $1\%$ to $5$     & $1\%$ to $10$  & $1\%$ to $15$     &       & $0\%$     & $1\%$ to $5$     & $1\%$ to $10$  & $1\%$ to $15$    &       & $0\%$     & $1\%$ to $5$     & $1\%$ to $10$  & $1\%$ to $15$ \\
        \cmidrule{1-1}\cmidrule{3-6}\cmidrule{8-11}\cmidrule{13-16}
    $n=1000$  &       & 0.90  & 0.88  & 0.87  & 0.85  &       & 0.91  & 0.89  & 0.87  & 0.85  &       & 0.90  & 0.88  & 0.87  & 0.86 \\
    $n=2500$  &       & 0.90  & 0.88  & 0.86  & 0.84  &       & 0.91  & 0.89  & 0.86  & 0.84  &       & 0.90  & 0.88  & 0.86  & 0.84 \\
    $n=5000$  &       & 0.90  & 0.88  & 0.86  & 0.84  &       & 0.91  & 0.89  & 0.87  & 0.83  &       & 0.90  & 0.88  & 0.86  & 0.84 \\
    $n=7500$  &       &       &       &       &       &       & 0.91  & 0.90  & 0.87  & 0.84  &       & 0.90  & 0.89  & 0.86  & 0.84 \\
    $n=10000$ &       &       &       &       &       &       & 0.91  & 0.90  & 0.87  & 0.83  &       & 0.90  & 0.88  & 0.86  & 0.83 \\
          &       &       &       &       &       &       &       &       &       &       &       &       &       &       &  \\
          &       & \multicolumn{4}{c}{EPP($\alpha=0.2$)}       &       & \multicolumn{4}{c}{EPP($\alpha=0.4$)}       &       & \multicolumn{4}{c}{EPP($\alpha=0.5$)} \\
    \multicolumn{1}{l}{Contamination} &       & $0\%$     & $1\%$ to $5$     & $1\%$ to $10$  & $1\%$ to $15$ &       & $0\%$     & $1\%$ to $5$     & $1\%$ to $10$  & $1\%$ to $15$     &       & $0\%$     & $1\%$ to $5$     & $1\%$ to $10$  & $1\%$ to $15$ \\
        \cmidrule{1-1}\cmidrule{3-6}\cmidrule{8-11}\cmidrule{13-16}
    $n=1000$  &       & 0.91  & 0.89  & 0.87  & 0.85  &       & 0.90  & 0.88  & 0.87  & 0.86  &       & 0.90  & 0.89  & 0.87  & 0.86 \\
    $n=2500$  &       & 0.90  & 0.89  & 0.86  & 0.84  &       & 0.90  & 0.88  & 0.86  & 0.85  &       & 0.90  & 0.88  & 0.86  & 0.85 \\
    $n=5000$  &       & 0.91  & 0.89  & 0.86  & 0.83  &       & 0.90  & 0.88  & 0.86  & 0.84  &       & 0.90  & 0.89  & 0.86  & 0.85 \\
    $n=7500$  &       & 0.91  & 0.89  & 0.86  & 0.83  &       & 0.90  & 0.89  & 0.86  & 0.84  &       & 0.90  & 0.89  & 0.86  & 0.84 \\
    $n=10000$ &       & 0.90  & 0.89  & 0.86  & 0.83  &       & 0.90  & 0.89  & 0.86  & 0.83  &       & 0.90  & 0.89  & 0.86  & 0.84 \\
          &       &       &       &       &       &       &       &       &       &       &       &       &       &       &  \\
          &       & \multicolumn{4}{c}{MREPP($L=2$)}     &       & \multicolumn{4}{c}{MREPP($L=4$)}     &       & \multicolumn{4}{c}{MREPP($L=6$)} \\
    \multicolumn{1}{l}{Contamination} &       & $0\%$     & $1\%$ to $5$     & $1\%$ to $10$  & $1\%$ to $15$ &       & $0\%$     & $1\%$ to $5$     & $1\%$ to $10$  & $1\%$ to $15$     &       & $0\%$     & $1\%$ to $5$     & $1\%$ to $10$  & $1\%$ to $15$ \\
        \cmidrule{1-1}\cmidrule{3-6}\cmidrule{8-11}\cmidrule{13-16}
    $n=1000$  &       & 0.92  & 0.91  & 0.90  & 0.88  &       & 0.91  & 0.91  & 0.89  & 0.88  &       & 0.91  & 0.90  & 0.89  & 0.88 \\
    $n=2500$  &       & 0.91  & 0.92  & 0.91  & 0.90  &       & 0.91  & 0.91  & 0.91  & 0.90  &       & 0.91  & 0.91  & 0.91  & 0.90 \\
    $n=5000$  &       & 0.91  & 0.92  & 0.92  & 0.92  &       & 0.91  & 0.91  & 0.92  & 0.91  &       & 0.91  & 0.91  & 0.92  & 0.91 \\
    $n=7500$  &       & 0.91  & 0.92  & 0.93  & 0.92  &       & 0.91  & 0.91  & 0.92  & 0.92  &       & 0.91  & 0.91  & 0.92  & 0.92 \\
    $n=10000$ &       & 0.90  & 0.91  & 0.92  & 0.92  &       & 0.91  & 0.91  & 0.92  & 0.92  &       & 0.91  & 0.91  & 0.92  & 0.92 \\
        \bottomrule
    \end{tabular}
    }
  \vspace{0.2cm}
\caption{Scenario 3: Prediction HPD $90\%$ coverage under increasing contamination. The empirical coverage is shown for increasing contamination level and sample sizes.}\label{tab:cov-cont}%
\end{table}%

\noindent\emph{Computational efficiency.}
Table~\ref{tab:time} reports the average runtime (in seconds). As expected, the runtime of MREPP increases with~$L$, whereas that of EPP decreases as $\alpha$ increases. Since the most computationally expensive component in both EPP and MREPP is the SPVT step, MREPP is consistently slower than EPP because it performs the SPVT $L$ times and additionally requires estimation of the resolution weights. Depending on the choice of $L$ and~$\alpha$, MREPP (EPP) is up to 4 (up to 36) times faster than the full GP when $n=5000$, and up to 4 (up to 47) times faster than LAGP when $n=10000$, while maintaining comparable predictive accuracy. Notably, EPP is the overall fastest prediction method\footnote{Note that MREPP can be parallelized across resolutions to achieve improved runtime.}. The PP is also competitive in terms of runtime, although its performance is more sensitive to the choice of the smoothness parameter~$\gamma$.
\begin{table}[t]
  \centering
  \resizebox{0.9\textwidth}{!}{
    \begin{tabular}{clrrrrrrrrrrrr}
    \toprule
          &       &       & \multicolumn{5}{c}{$\nu=0.5$ (Exponential)}                &       & \multicolumn{5}{c}{$\nu=1.5$ (Mat\'{e}rn)} \\
              \cmidrule{4-8}\cmidrule{10-14}
        & $n$     &       & $1000$     & $2500$     & $5000$     & $7500$ & $10000$     &       & $1000$     & $2500$     & $5000$     & $7500$ & $10000$ \\
            \cmidrule{1-2}\cmidrule{4-8}\cmidrule{10-14}
            \multirow{9}[0]{*}{Fixed Space $[-3,3]^3$} 
        & GP    &   & 0.7 & 9.3 & 70.5 &  229.1$^\ast$ & 539.5$^\ast$  &   & 0.6 & 7.6 & 65.1 &  231.1$^\ast$ & 523.0$^\ast$  \\
        & PP    &   & 1.8   & 10.2  & 44.8  & 113.0 & 223.8 &       & 1.8   & 5.3   & 9.7   & 15.6  & 22.6 \\
        & LAGP  &   & 13.4  & 20.1  & 38.7  & 108.5 & 165.3 &       & 13.8  & 20.7  & 40.5  & 110.6 & 159.3 \\
        & EPP($\alpha=0.2$)   &   & 1.5   & 3.6   & 9.5   & 16.7  & 28.7  &       & 1.5   & 3.1   & 4.6   & 6.3   & 8.4 \\
        & EPP($\alpha=0.4$)   &   & 0.6   & 0.8   & 1.3   & 5.6   & 14.2  &       & 0.6   & 0.8   & 1.3   & 4.9   & 11.1 \\
        & EPP($\alpha=0.5$)   &   & 0.7   & 0.9   & 1.6   & 2.4   & 3.5   &       & 0.7   & 0.9   & 1.6   & 2.4   & 3.4 \\
        & MREPP($L=2$) &   & 3.7   & 5.6   & 8.8   & 11.8  & 15.2  &       & 3.7   & 5.7   & 8.8   & 12.0  & 15.1 \\
        & MREPP($L=4$) &   & 6.2   & 11.0  & 16.8  & 27.0  & 40.7  &       & 6.3   & 10.5  & 16.3  & 25.8  & 37.2 \\
        & MREPP($L=6$) &   & 8.3   & 16.8  & 27.9  & 43.6  & 61.8  &       & 8.3   & 16.0  & 25.4  & 38.0  & 52.6 \\
        \midrule
    \multirow{9}[0]{*}{Fixed Radius (0.001)} 
        & GP    &   & 0.6 & 7.1 & 61.5 & 225.1$^\ast$ & 527.7$^\ast$ &   & 0.6 & 7.1 & 61.3 &  227.1$^\ast$ & 510.7$^\ast$  \\
        & PP    &   & 1.8   & 10.2  & 44.6  & 112.0 & 223.2 &       & 0.6   & 2.5   & 11.8  & 44.3  & 122.8 \\
        & LAGP  &   & 13.4  & 20.0  & 38.6  & 109.5 & 166.1 &       & 13.8  & 20.7  & 40.3  & 110.4 & 158.1 \\
        & EPP($\alpha=0.2$)   &   & 1.5   & 3.6   & 9.5   & 16.8  & 28.8  &       & 1.3   & 2.5   & 5.1   & 11.6  & 23.9 \\
        & EPP($\alpha=0.4$)   &   & 0.6   & 0.8   & 1.3   & 5.6   & 14.3  &       & 0.6   & 0.8   & 1.3   & 5.4   & 13.8 \\
        & EPP($\alpha=0.5$)   &   & 0.7   & 0.9   & 1.6   & 2.4   & 3.5   &       & 0.7   & 0.9   & 1.6   & 2.4   & 3.4 \\
        & MREPP($L=2$) &   & 3.7   & 5.6   & 8.8   & 11.8  & 15.2  &       & 3.4   & 5.7   & 8.8   & 12.0  & 14.9 \\
        & MREPP($L=4$) &   & 6.2   & 11.0  & 16.8  & 26.9  & 40.6  &       & 5.7   & 9.8   & 16.7  & 26.8  & 39.9 \\
        & MREPP($L=6$) &   & 8.3   & 16.7  & 27.9  & 43.7  & 61.8  &       & 7.8   & 15.0  & 26.1  & 42.0  & 60.2 \\
          \bottomrule
    \end{tabular}%
  }
  \vspace{0.2cm}
\caption{Scenarios 1 and 2: Average runtime across replications in seconds. $^\ast$The GP performance for $n>5000$ is based on $10$ replicates for computational reasons.}\label{tab:time}%
\end{table}

\paragraph{Overall summary.} In settings without contamination, the EPP and MREPP achieve prediction accuracy comparable to the full GP and LAGP while being substantially faster. Furthermore,   both approaches outperform the PP, particularly when  $\gamma < 2$. However, the MREPP yields superior point and probabilistic predictions with more accurate uncertainty quantification than EPP under training data contamination. This is because the MREPP is capable of adapting effectively to training data contamination by shifting weight from finer to coarser resolutions. To conclude, the MREPP offers clear practical advantages over existing approaches by providing robustness to contamination without sacrificing accuracy in standard settings while remaining computationally scalable to high-dimensional data.

\subsection{Real data illustrations}
\label{sec:appl}
To illustrate the efficacy of MREPP compared to the benchmark methods considered in Section \ref{sec:sim} on five real datasets of varying sample size and spatial distributions, see the Supplementary Material F for their visualizations. 

\paragraph{Datasets.} We consider the following five real datasets that differ in size and characteristics:
\begin{enumerate}
    \item {\it Brain} dataset from the \textsf{R}-package \texttt{gamair}. It contains lattice data of one slice of  functional magnetic resonance imaging measurements for a human brain. The total number of observations is $1567$ which we divide into $n=1000$ and $N=567$ for training and testing, respectively. Since this dataset contains outliers, we use it to evaluate the predictive robustness by allocating values greater than the $99\%$ quantile to the training set with high probability.
    \item {\it Precipitation} dataset contains total precipitations in mm during 2023 over a region of the Pacific Ocean between longitude $[-150,-70]$ and latitude $[-30,30]$. These data are downloaded from  \textsf{R}-package \texttt{nasapower}. The total number of observations is $3483$ which we divide into $n=3000$ and $N=483$ for training and testing, respectively. Since this dataset contains outliers, we use it to evaluate the predictive robustness by allocating values greater than the $995\%$ quantile to the training set with high probability.
    \item {\it Anomalies} dataset used in \cite{SonDaiGen2025}. It consists of spatially non-uniformly distributed annual total precipitation anomalies observed in the United States in 1962 at $7352$ weather stations. We randomly divide the data into $n=7000$ training  and $N = 352$ test locations. 
    \item {\it Ozone} dataset used in \cite{SonDaiGen2025}. It provides the level-2 total column ozone for October 1, 1988 at $173,405$ locations over a regularly shaped region, and we randomly divide the data into $n=150,000$ training observations and $N = 23,405$ test locations.
    \item {\it Canopy} dataset used in \cite{CooCorNel2013} and \cite{MaoMarRei2024}. It collects $1,723,137$ measurements of canopy height in the Connecticut River Valley. We randomly divide the total observations into $n=1,378,510$ ($80\%$ of the total data points) and $N=344,627$ for training and testing, respectively.
\end{enumerate}

\paragraph{Hyperparameter settings.}
For the first four datasets {\it Brain}, {\it Precipitation}, {\it Anomalies} and {\it Ozone}, we consider the hyperparameter settings and the spatial partitioning scheme used in Section \ref{sec:sim} for EPP and MREPP. Since the {\it Canopy} dataset is  high-dimensional,  we consider $\alpha \in \{0.4, 0.5, 0.6\}$ (i.e., we replace $\alpha=0.2$ with $\alpha=0.6$ for computational limits) for EPP, and $\alpha_L=0.6$ for the last resolution of MREPP. Following Section \ref{sec:tune-MREPP}, we use $m_{\max}=20$ for {\it Brain} and {\it Precipitation}, $m_{\max}=100$ for {\it Anomalies}, and $m_{\max}=200$ for {\it Ozone} and {\it Canopy}, respectively. 

LAGP is implemented using the ALC algorithm to construct the local process, following the approach in Section \ref{sec:sim}, for the first three datasets. For {\it Canopy}, however, we instead use only the five nearest neighbors, due to computational constraints.

The PP is estimated using $m=500$ for {\it Brain} and $m=1000$ for {\it Precipitation}, and, following \cite{SonDaiGen2025}, $m=1000$ and $m=1755$ for {\it Anomalies} and {\it Ozone}, respectively. Finally, we use $m=5000$ for {\it Canopy} because of computational constraints. For computational reasons, we do not consider the full GP for the {\it Ozone} and {\it Canopy} datasets. 

For convenience, in a first step, we center the observations around their empirical mean for each dataset. Then, we estimate a zero-mean GP using the function \texttt{fit\_model} in the \texttt{GpGp} package. This function employs a  Vecchia approximation. We then fix the Mat\'{e}rn kernel and nugget effect for each benchmark method equal to the ones estimated with \texttt{fit\_model} to ensure comparability. Details on these estimates are given in Table F.1 in the Supplementary Material.

\paragraph{Results.} Table \ref{tab:resappl} summarizes  point and probabilistic prediction accuracy (MSE and LPS), uncertainty quantification through predictive $90\%$ HPD interval coverage, and  runtime (in  minutes), for the {\it Anomalies}, {\it Ozone}, and {\it Canopy} datasets. Table \ref{tab:resrob} shows the performance of methods with respect to predictive robustness assessed in the {\it Brain} and {\it Precipitation} datasets.

\noindent\emph{Point and probabilistic prediction}. Across the three datasets  {\it Anomalies}, {\it Ozone}, and {\it Canopy}, the proposed EPP and MREPP  achieve MSE that is competitive with the benchmark methods, with MREPP attaining the best performance for the {\it Canopy} dataset. In particular, MREPP achieves a point prediction performance equivalent to the best EPP, i.e., EPP($\alpha_3$), as, in the absence of outliers, its resolution weights correctly focus on finer resolutions, thus favoring prediction accuracy over robustness. This empirical finding is in line with theoretical results for MREPP. While the performance of LAGP on {\it Canopy} could potentially be improved by increasing the number of nearest neighbors or by optimizing the local design using ALC, such improvements would come at substantially higher computational cost. In terms of probabilistic prediction,  MREPP consistently delivers the best LPS across all three datasets, with a performance increase for a higher number of resolutions.

\noindent\emph{Uncertainty quantification}. For the {\it Anomalies} dataset, all methods achieve  coverage close to the $90\%$ level, indicating that predictive uncertainty is well calibrated. In contrast, for both the {\it Ozone} and {\it Canopy} datasets, all methods exhibit similar behavior in overestimating coverage.

\noindent\emph{Computational efficiency}. We find strongly favorable results for EPP and MREPP in terms of computational cost relative to GP, PP, and LAGP. In the {\it Anomalies} dataset, runtime decreases from 4.53 minutes for GP to well below one minute for both EPP and MREPP. As the size of the dataset increases, advantages of EPP and MREPP become more evident. For {\it Ozone}, EPP and MREPP are approximately nine times faster than LAGP, depending on $\alpha$ and $L$. PP is competitive in this dataset, but it provides much worse predictive performance. The most pronounced improvements appear in the {\it Canopy} dataset, where PP requires nearly 47 minutes and LAGP about 18 minutes, while EPP($\alpha_3=0.6$) and MREPP($L=6$) are about 19 and 2.3 times faster than LAGP, respectively.

\begin{table}[htbp]
  \centering
    \resizebox{1\textwidth}{!}{
    \begin{tabular}{lrrrrrrrrrrrrrrr}
    \toprule
          &       & \multicolumn{4}{c}{\it Anomalies} &       & \multicolumn{4}{c}{\it Ozone}     &       & \multicolumn{4}{c}{\it Canopy} \\
          &       & \multicolumn{1}{c}{MSE} & \multicolumn{1}{c}{LPS} & \multicolumn{1}{c}{Coverage} & \multicolumn{1}{c}{Runtime} &       & \multicolumn{1}{c}{MSE} & \multicolumn{1}{c}{LPS} & \multicolumn{1}{c}{Coverage} & \multicolumn{1}{c}{Runtime} &       & \multicolumn{1}{c}{MSE} & \multicolumn{1}{c}{LPS} & \multicolumn{1}{c}{Coverage} & \multicolumn{1}{c}{Runtime} \\
          \cmidrule{1-1}\cmidrule{3-6}\cmidrule{8-11}\cmidrule{13-16}
    GP    &       & 0,22  & 0,66  & 0,90  & 4,53  &       &       &       &       &       &       &       &       &       &  \\
    PP    &       & 0,25  & 0,74  & 0,92  & 0,32  &       & 37,59 & 3,86  & 0,94  & 2,56  &       & 7,69  & 2,68  & 0,96  & 46,92 \\
    LAGP  &       & 0,22  & 0,65  & 0,90  & 2,92  &       & 28,37 & 4,22  & 0,94  & 9,26  &       & 1,71  & 3,00  & 0,94  & 17,73 \\
    EPP($\alpha_1$)  &       & 0,24  & 0,73  & 0,90  & 0,30  &       & 27,71 & 4,12  & 0,94  & 1,12  &       & 1,83  & 2,99  & 0,94  & 21,42 \\
    EPP($\alpha_2$)  &       & 0,23  & 0,68  & 0,91  & 0,15  &       & 27,48 & 4,12  & 0,94  & 0,81  &       & 1,96  & 2,68  & 0,95  & 3,60 \\
    EPP($\alpha_3$)  &       & 0,22  & 0,66  & 0,90  & 0,12  &       & 27,48 & 4,10  & 0,94  & 0,56  &       & 1,66  & 2,86  & 0,94  & 0,93 \\
    MREPP($L=2$) &       & 0,22  & 0,65  & 0,91  & 0,18  &       & 27,58 & 3,98  & 0,94  & 0,64  &       & 1,66  & 2,85  & 0,94  & 1,73 \\
    MREPP($L=4$) &       & 0,22  & 0,65  & 0,91  & 0,41  &       & 27,58 & 3,94  & 0,94  & 0,98  &       & 1,66  & 2,73  & 0,95  & 4,15 \\
    MREPP($L=6$) &       & 0,22  & 0,65  & 0,91  & 0,66  &       & 27,58 & 3,90  & 0,94  & 1,45  &       & 1,66  & 2,56  & 0,95  & 7,70 \\
    \bottomrule
    \end{tabular}
    }
      \vspace{0.2cm}
\caption{Real data: Point and probabilistic predictions, uncertainty quantification, and runtime. The MSE, LPS, the predictive $90\%$ HPD coverage, and the runtime in minutes are shown for {\it Anomalies}, {\it Ozone}, and {\it Canopy} data. For EPP, $(\alpha_1,\alpha_2,\alpha_3)=(0.2,0.4,0.5)$ for {\it Anomalies} and {\it Ozone}, and $(\alpha_1,\alpha_2,\alpha_3)=(0.4,0.5,0.6)$ for {\it Canopy}. For the high-dimensional datasets {\it Ozone} and {\it Canopy} computations are parallelized across 20 CPU cores.}
  \label{tab:resappl}%
\end{table}%

\noindent\emph{Robustness}. The results in Table \ref{tab:resrob} highlight the predictive robustness of MREPP to outliers in the training set. The overall better performance of MREPP can be more clearly understood when the metrics are evaluated separately on the nearest neighbors of each outlier and on the remaining locations. In regions far from outliers, all methods perform similarly across MSE, LPS, and coverage. However, greater differences arise in neighborhoods close to anomalous observations. The GP, PP, LAGP, and EPP exhibit a degradation of the predictive performance near outliers and severe undercoverage of predictive intervals. The poor performance of PP can be explained by the choice of large $m=500$ compared to $n$. In fact, using a much smaller value could improve its behavior, albeit at the expense of global accuracy. In contrast, MREPP delivers improvements in both point and density prediction near outliers, while maintaining good coverage above $85\%$ for {\it Brain} and around $75\%$ for {\it Precipitation}. This behavior can be attributed to the resolution-weighting mechanism, which assigns importance also to coarser resolutions where the number of SPs remains limited, thereby preserving robustness. Notably, although the resolution weights are estimated by minimizing the MSE on a calibration set (i.e., using point prediction alone), the gains of MREPP under training data contamination carry over to probabilistic prediction.

\begin{table}[htbp]
  \centering
  \resizebox{1.0\textwidth}{!}{
    \begin{tabular}{llrrrrrrrrrrrr}
    \toprule
          &       &       & \multicolumn{3}{c}{Overall} &       & \multicolumn{3}{c}{Near outliers} &       & \multicolumn{3}{c}{Far from outliers} \\
          \cmidrule{4-6}\cmidrule{8-10}\cmidrule{12-14}
          &       &       & \multicolumn{1}{c}{MSE} & \multicolumn{1}{c}{LPS} & \multicolumn{1}{c}{Coverage} &       & \multicolumn{1}{c}{MSE} & \multicolumn{1}{c}{LPS} & \multicolumn{1}{c}{Coverage} &       & \multicolumn{1}{c}{MSE} & \multicolumn{1}{c}{LPS} & \multicolumn{1}{c}{Coverage} \\
          \cmidrule{1-2}\cmidrule{4-6}\cmidrule{8-10}\cmidrule{12-14}
        \multirow{9}{*}{{\it Brain}}
          & GP    &       & 1,45  & 2,15  & 0,92  &       & 8,72  & 11,06 & 0,50  &       & 0,83  & 1,39  & 0,96 \\
          & PP    &       & 1,60  & 2,30  & 0,92  &       & 9,91  & 12,41 & 0,53  &       & 0,90  & 1,44  & 0,95 \\
          & LAGP  &       & 1,46  & 2,14  & 0,92  &       & 8,70  & 10,93 & 0,50  &       & 0,83  & 1,39  & 0,96 \\
          & EPP($\alpha=0.2$)   &       & 1,50  & 2,19  & 0,92  &       & 9,01  & 11,22 & 0,56  &       & 0,87  & 1,43  & 0,96 \\
          & EPP($\alpha=0.4$)   &       & 1,43  & 2,12  & 0,92  &       & 8,51  & 10,64 & 0,54  &       & 0,83  & 1,40  & 0,96 \\
          & EPP($\alpha=0.5$)   &       & 1,41  & 2,09  & 0,93  &       & 8,29  & 10,26 & 0,56  &       & 0,83  & 1,40  & 0,96 \\
          & MREPP($L=2$) &       & 1,04  & 1,53  & 0,97  &       & 3,26  & 1,97  & 0,94  &       & 0,85  & 1,49  & 0,97 \\
          & MREPP($L=4$) &       & 1,25  & 1,60  & 0,96  &       & 5,95  & 3,10  & 0,86  &       & 0,85  & 1,48  & 0,97 \\
          & MREPP($L=6$) &       & 1,23  & 1,58  & 0,97  &       & 5,81  & 2,82  & 0,87  &       & 0,84  & 1,47  & 0,97 \\
    \midrule
    \multirow{9}{*}{{\it Precipitation}}
          & GP    &       & 0,019 & 3,40  & 0,91  &       & 0,252 & 77,46 & 0,29  &       & 0,004 & -1,20 & 0,94 \\
          & PP    &       & 0,019 & 2,64  & 0,90  &       & 0,231 & 53,61 & 0,30  &       & 0,006 & -0,52 & 0,94 \\
          & LAGP  &       & 0,018 & 3,37  & 0,91  &       & 0,250 & 76,78 & 0,29  &       & 0,004 & -1,20 & 0,94 \\
          & EPP($\alpha=0.2$)   &       & 0,018 & 1,50  & 0,91  &       & 0,219 & 40,07 & 0,35  &       & 0,006 & -0,89 & 0,95 \\
          & EPP($\alpha=0.4$)   &       & 0,018 & 2,85  & 0,91  &       & 0,243 & 67,88 & 0,36  &       & 0,004 & -1,19 & 0,95 \\
          & EPP($\alpha=0.5$)   &       & 0,018 & 2,57  & 0,92  &       & 0,244 & 63,01 & 0,38  &       & 0,004 & -1,18 & 0,95 \\
          & MREPP($L=2$) &       & 0,018 & -0,29 & 0,95  &       & 0,176 & 8,06  & 0,59  &       & 0,008 & -0,81 & 0,97 \\
          & MREPP($L=4$) &       & 0,017 & -0,56 & 0,97  &       & 0,169 & 3,17  & 0,76  &       & 0,007 & -0,79 & 0,98 \\
          & MREPP($L=6$) &       & 0,017 & -0,62 & 0,97  &       & 0,177 & 2,16  & 0,75  &       & 0,007 & -0,79 & 0,99 \\
    \bottomrule
    \end{tabular}
    }
      \vspace{0.2cm}
\caption{{\it Brain} and {\it Precipitation} data: Point and probabilistic predictions, and uncertainty quantification in presence of outliers. The MSE, LPS, and the predictive $90\%$ HPD coverage are shown for the {\it Brain} and {\it Precipitation} datasets and computed on different test sets: ``Overall" considers all  test locations; ``Near outliers" considers the set of five nearest neighbor locations to each outlier; ``Far from outliers"  is obtained as ``Overall" minus ``Near outliers".}
  \label{tab:resrob}%
\end{table}%

\paragraph{Overall summary.} The empirical findings indicate that MREPP achieves point prediction accuracy comparable to or better than the benchmark methods and consistently outperforms the latter in terms of probabilistic prediction, while preserving low computational cost also in very high-dimensional datasets. Furthermore, MREPP maintains good prediction performance and reliable coverage also in the presence of data contamination in the training set. These results make MREPP an effective method to balance statistical and computational efficiency in possibly large-scale spatial prediction problems.

\section{Discussion}
\label{sec:concl}
This paper addresses a central challenge in spatial statistics: achieving accurate and  robust predictions while being scalable to high-dimensional data.

We address this challenge with three contributions. First, we introduce the \emph{Ensemble of Predictive Processes} (EPP) together with  a spatial partitioning algorithm and patching scheme, and we establish convergence rates for the EPP, thereby quantifying the trade-off between statistical accuracy and computational scalability. Second, we demonstrate that PPs with fixed inducing points are more robust to training data contamination than full GPs and the EPP. Third, we introduce the MREPP, a method that integrates recent advances in PPs with divide-and-conquer strategies across multiple spatial resolutions. Using coarse-to-fine spatial partitions, the MREPP is computationally efficient, balances predictive robustness with local adaptivity, and delivers accurate predictions even in the presence of training data contamination. Overall, MREPP provides a practical and theoretically grounded solution for modern large-scale spatial modeling  in the context of scalable and more robust GP prediction. 

Empirical results on synthetic and real-world datasets confirm that the MREPP achieves  predictive accuracy and uncertainty quantification comparable to those of full GPs, while being orders of magnitude faster. Moreover, the MREPP adapts to training data contamination by shifting weights toward coarser resolutions, and consequently outperforms state-of-the-art global and local approximations in terms of predictive robustness.

Although this paper focuses on prediction, the process parameters and nugget variance could, in principle, be estimated by maximizing the likelihood of MREPP. As an advantage, the MREPP naturally enables local, resolution-specific parameter estimation by restricting inference to observations within each spatial partition. Unlike standard PPs with globally shared parameters, this capability may make the MREPP beneficial  for flexibly accommodating nonstationary spatial structures. We consider this aspect beyond the scope of this paper and leave it for future research. 

Beyond the aspects mentioned, there are other interesting future directions.  First,  extending the methodology to allow resolution probabilities to vary with location $p(l|s_\star)$ could improve flexibility in heterogeneous domains and clustered anomalies \citep{WanSonWan2024}. 
Second, tailoring the MREPP to spatio-temporal prediction constitutes a natural next step which would require non-trivial construction of dynamic partitions and time-adaptive inducing sets. Third, although our notion of predictive robustness focuses on how training data contamination affects the predictive mean, future work could investigate the robustness of the entire predictive distribution \citep{AltBriKno2024}.   
Finally, characterizing  the MREPP from the direction of probabilistic numerics \citep{StaSza2024} could broaden its theoretical understanding.



\bibliographystyle{apalike}
\bibliography{biblio}
	
\end{document}